\documentclass[usenatbib]{mn2e}
\usepackage{threeparttable,multirow}

\usepackage{amsmath,amssymb}

\usepackage{graphicx}
\usepackage{subfigure}
\usepackage{color}
\usepackage{times}

\usepackage{url}

\def\spose#1{\hbox to 0pt{#1\hss}}
\newcommand{\lta}{\mathrel{\spose{\lower 3pt\hbox{$\mathchar"218$}}
    \raise 2.0pt\hbox{$\mathchar"13C$}}}
\newcommand{\gta}{\mathrel{\spose{\lower 3pt\hbox{$\mathchar"218$}}
    \raise 2.0pt\hbox{$\mathchar"13E$}}}

\mathchardef\star="313F

\newcommand{\model}{{\cal M}}
\newcommand{\data} {{\bf D}}
\newcommand{\param}{\boldsymbol{\theta}}

\newcommand{\tup}{\hat T}
\newcommand{\tdown}{\check T}

\newcommand{\mnras}{MNRAS}   % Monthly Notices
\begin{document}

\title[Bayesian Inference Engine]{Computational statistics using the
  Bayesian Inference Engine}
\author[]{Martin D. Weinberg$^{1}$ \\
  $^1$ Department of Astronomy, University of Massachusetts, Amherst
  MA 01003-9305 }

\date{}

\pagerange{\pageref{firstpage}--\pageref{lastpage}}
\pubyear{2013}

\maketitle

\label{firstpage}

%%%%%%%%%%%%%%%%%%%%%%%%%%%%%%%%%%%%%%%%%%%%%%%%%%%%%%%%%%%%%%%%%%%%%%%%%%
\begin{abstract}
  This paper introduces the Bayesian Inference Engine (BIE), a general
  parallel, optimised software package for parameter inference and
  model selection.  This package is motivated by the analysis needs of
  modern astronomical surveys and the need to organise and reuse
  expensive derived data.  The BIE is the first platform for
  computational statistics designed explicitly to enable Bayesian
  update and model comparison for astronomical problems.  Bayesian
  update is based on the representation of high-dimensional posterior
  distributions using metric-ball-tree based kernel density
  estimation.  Among its algorithmic offerings, the BIE emphasises
  hybrid tempered MCMC schemes that robustly sample multimodal
  posterior distributions in high-dimensional parameter spaces.
  Moreover, the BIE is implements a full \emph{persistence} or
  \emph{serialisation} system that stores the full byte-level image of
  the running inference and previously characterised posterior
  distributions for later use.  Two new algorithms to compute the
  marginal likelihood from the posterior distribution, developed for
  and implemented in the BIE, enable model comparison for complex
  models and data sets.  Finally, the BIE was designed to be a
  collaborative platform for applying Bayesian methodology to
  astronomy.  It includes an extensible object-oriented and easily
  extended framework that implements every aspect of the Bayesian
  inference.  By providing a variety of statistical algorithms for all
  phases of the inference problem, a scientist may explore a variety
  of approaches with a single model and data implementation.
  Additional technical details and download details are available from
  \url{http://www.astro.umass.edu/bie}.  The BIE is distributed under
  the GNU GPL.
\end{abstract}
%%%%%%%%%%%%%%%%%%%%%%%%%%%%%%%%%%%%%%%%%%%%%%%%%%%%%%%%%%%%%%%%%%%%%%%%%%


%%%%%%%%%%%%%%%%%%%%%%%%%%%%%%%%%%%%%%%%%%%%%%%%%%%%%%%%%%%%%%%%%%%%%%%%%%

\section{Introduction}
\label{sec:intro}

Inference is fundamental to the scientific process.  We may broadly
identify two categories of inference problems: 1) {\em
  estimation}---finding the parameter of a theory or model from data;
and 2) {\em hypothesis testing}---determining which theory, indeed if
any, is supported by the data.  Astronomers increasingly rely on
numerical data analysis, but most cannot take full advantage of the
power afforded by present-day computational statistics for attacking
the inference problem owing to a lack of tools.  This is especially
critical when data comes from multiple instruments and surveys.  The
different data characteristics of each survey include varied
selection effects and inhomogeneous error models.  Moreover, the
information content of large survey databases can in principle
determine models with many parameters but exhaustive exploration of
parameter space is often not feasible.

These classes of estimation problems are readily posed by Bayesian
inference, which determines model parameters, $\param$, while allowing
for straightforward incorporation of heterogeneous selection biases.
In the Bayesian paradigm, current knowledge about the model parameters
is expressed as a probability distribution called the {\em prior
  distribution}, $P(\param)$. This is the anticipated distribution
of parameters for the postulated model \emph{before} obtaining any
measurements. This should include one's understanding of the model
parameters in their theoretical context. When new data $\data$ becomes
available, the information content is expressed as $P(\data|\param)$,
the distribution of the observed data given the model parameters.
This will be familiar to some as the classical {\em likelihood}
function, $L(\data|\param)$.  This information is then combined with
the prior to produce an updated probability distribution called the
{\em posterior distribution}, $P(\param|\data)$.  Bayes' Theorem
defines this update mathematically:
\begin{equation}
  P(\param|\data) = \frac{P(\param)P(\data|\param)}{\int P(\param)
    P(\data|\param)\,d\param}.
  \label{eq:Bayes}
\end{equation}
Equation (\ref{eq:Bayes}) is a simply application of the
multiplicative rule for conditional probability.  Combined with the
concept of sample spaces, measure theory, and Monte Carlo computation,
Bayes theorem provides a rich framework for the quantitative
investigation of a wide variety of inference problems, such as
classification and cluster analyses, which broadly extends the two
groups described above.  Later sections will illustrate the importance
and utility of explicit quantification of the prior information. More
generally, equation (\ref{eq:Bayes}) emphasises that Bayesian
inference is about probability \emph{distributions}; we will see below
that the BIE is, essentially, a computational tool that manipulates
probability distributions as objects.

Astronomy is a data-rich subject, and methods of mathematical
statistics have been applied to every branch astronomical research.
The BIE was designed to implement a solution to computational solution
to inference based on parametric models on very large data sets, in
particular.  For other topics in astrostatitics,
\citet{Babu.Feigelson:96} present a comprehensive overview of
nonparametric methods, multivariate analysis, time series analysis,
density estimation, and resampling methods.  More recently, motivated
by new computational approaches and fast computers just as the BIE,
Bayesian approaches are now mainstream and several extensive
monographs and textbooks emphasising the Bayesian approach are now
available.  To name a few, \citet{Gregory:2005} presents many of these
same applications described in \citet{Babu.Feigelson:96} from the
Bayesian point of view and provides many useful worked examples.  For
observational astronomy per se, \citet{Wall.Jenkins:2012} nicely
describes Bayesian approaches to inferring galaxy luminosity functions
and spatial correlation functions.  Finally, \citet{Hobson.etal:2010}
introduces and reviews the use of Bayesian methods in cosmological
data analysis problems, describing approaches to source detection,
galaxy population analysis and classification and the inference of
cosmological parameters from cosmic microwave background.

Why use the Bayesian framework? To begin, the Bayesian approach
unifies both aspects of the inference problem: estimation and
hypothesis testing.  For example, given a galaxy image and several
families of brightness profiles, we would like to determine both the
distribution of parameters for each family and which family is best
supported by the data.  A classical analysis might report the Maximum
Likelihood (ML) estimate for each model using a $\chi^2$-type
statistic \citep{Pearson:1900} and prefer the fit with the lowest
value of $\chi^2$ per degree of freedom.  However, the $\chi^2$
statistic will grow with sample size in the presence of measurement
errors.  This leads to the well-known over-fitting problem where the
Pearson-type $\chi^2$ test will reject the correct distribution in
favour of one which better describes the deviations caused by the
measurement errors.  This well-known issue may be treated in a variety
of ways, but the Bayesian approach naturally \emph{prefers} the model
with the smallest number of dimensions that can explain the data
distribution through the specification of the prior information.  The
Bayesian approach further emphasises that model comparison problems
must depend on the prior distribution.  See Section \ref{sec:goodness}
for more details.

The computational complexity for a direct evaluation of equation
(\ref{eq:Bayes}) directly grows exponentially with the number of model
parameters and becomes intractable before the volume of currently
available large data sets is reached.  However, Monte Carlo algorithms
based on Markov chains for drawing samples from the posterior
distribution promise to make the Bayesian approach very widely
applicable \citep[e.g. see][]{robert.casella:04}.  In turn, the
application of Bayesian methods in all fields of astrophysics from
planetary detection to cosmology has been enabled by fast computers
and spurred by data sets of increasing size and complexity.  Once a
scientist can determine the posterior distribution, rigorous credible
bounds on parameters and powerful probability-based methods for
selecting between competing models and hypotheses immediately follow.
This statistical approach is superior to those commonly used in
astronomy because it makes more efficient use of all the available
information and allows one to test astronomical hypotheses directly.
To realise this promise for astronomical applications, we need a
software system designed to handle both large data sets and large
model spaces simultaneously.

Beginning in 2000, a multidisciplinary investigator team from the
Departments of Astronomy and Computer Science at UMass designed and
implemented the Bayesian Inference Engine\footnote{See
  \url{http://www.astro.umass.edu/bie} for detailed description and
  download instructions.}, a parallel software platform for performing
statistical inference over very large data sets.  We focus on
probability-based Bayesian statistical methods because they provide
maximum flexibility in incorporating and using all available
information in a model-data comparison.  For example, multiple data
sources can be naturally combined and their selection effects, which
must be specified by the data provider to obtain a meaningful
statistical inference, are easily incorporated.  In this way, the BIE
provides a platform for investigating inference using the virtual
observatory paradigm.

I begin in Section \ref{sec:wants_needs} by introducing the concepts
in Bayesian inference that illustrate its power as a framework for
parameter estimation and model selection for astronomical
problems. This power, not surprisingly, comes with significant
computational challenges that informed our design for the BIE.
Indeed, some of the most attractive features of Bayesian inference,
such as Bayesian update (see Section \ref{sec:bayes_update}) and
general non-nested model selection (see Section \ref{sec:modelsel}),
are inaccessible to many researchers owing the computational
complexity of dealing with distributions as objects.  This is provided
by the BIE intrinsically.  An additional major feature of the BIE is
ability to full save or \emph{persist} its full running state to disc.
Since the probability distributions are first-class objects in the
BIE, any of these may be recalled and reused at a later date.  All of
these features together enable a unique workflow.  The key features of
the package and the BIE-enabled workflow are described in Section
\ref{sec:design} and in detail in Section \ref{sec:software}.  This
is followed by a brief summary of BIE-enabled research in Section
{\ref{sec:examples}} as case studies.  I summarise in Section
\ref{sec:summary}.  Some of the key implementation details are
described in the Appendix.  In particular, Section \ref{sec:bie}
describes and motivates our choice of advanced MCMC algorithms
(Section \ref{sec:bie}).  Some of these were developed specifically
for high-dimensional astronomical research problems using the BIE.
Moreover, the BIE allows additional algorithms to be straightforwardly
added as needed (see Section \ref{sec:arch}).

\section{What do astronomers want and need?}
\label{sec:wants_needs}

\subsection{Parameter estimation}
\label{sec:paramest}

Many astronomical data analysis problems are posed as parameter
estimates.  For example: 1) estimate the temperature of an object from
its spectral energy distribution; or 2) estimate a galaxy's scale
length from its flux profile.  In these problems, one is asserting
that the underlying model is true and testing the hypothesis that the
parameter, temperature or scale length, has a particular value.

Bayesian inference approaches these problems with the following three
steps, reflecting the standard practice of the scientific method: 1)
numerically quantify a prior belief in the hypothesis; 2) collect data
that will either be consistent or inconsistent with the hypothesis; 3)
compute the new belief in the hypothesis given the new data.  These
steps may be repeated to achieve the desired degree of belief.  A
clever observer will design campaigns that refine the degree of belief
efficiently (i.e., that makes the belief in the hypothesis high or
low).  In the context of our simple examples, one may believe that the
spectral energy distribution is that of an M-dwarf star and one's
prior belief is then a distribution of values centred on 2000 K.
After measuring the spectral energy distribution, the prior
distribution of temperature is combined with the probability of
observing the data for a particular temperature, to get a refined
distribution of the temperature of the object.  Notice that this
procedure does not result in a single value.  Rather, the posterior
probability distribution is used to estimate a credible interval (also
known as a Bayesian confidence interval).  Of course, credible
intervals and regions are only a simple summary of the information
contained in the posterior distribution. Unlike classical statistics,
Bayesian inference does not rely on a significance evaluation based on
theoretical or empirical reference distributions that are valid in the
limit of very large data sets.  Rather it specifies the probability
distribution function for the parameters explicitly based on the data
at hand.

A prime motivation for the BIE project is the thesis that the power of
expensive and large survey data sets is underutilised by targeting
parameter estimation as the goal.  To illustrate this, let us consider
the second example above: estimating the scale length of a disc.  A
standard astronomical analysis might proceed as follows. One
determines the posterior probability distribution for scale lengths
for some subset of survey images.  Alongside scale length, one
determines other parameters such as luminosity, axis ratios, or
inclinations, and possibly higher moments such as the asymmetry.  The
scale length with maximum probability becomes the \emph{best estimate}
and is subsequently correlated with some other parameter of interest,
luminosity or asymmetry, say.  Then, any correlation is interpreted in
the context of theories of galaxy formation and evolution.  Observe,
that in the first step, one is throwing out much of the information
implicit in the posterior distribution.  In particular, the luminosity
estimate is most likely correlated with the scale-length estimate.  If
one were to plot the posterior distribution in these two parameters,
one might find that the distribution is elongated in the
scale-length--asymmetry plane, possibly in the same sense as the
putative correlation!  In other words, the confidence in the
hypothesis of a correlation should include the full posterior
distribution of parameter estimates, not just the maximum probability
estimate.  See Section \ref{sec:galphat}, Figure \ref{fig:galphat1} for a
real-world example.

Moreover, this scenario suggests that one is using disk scale length
and asymmetry as a proxy for testing a hypothesis about disk evolution
or environment.  These results might have been more reliable if the
observational campaign had been designed to enable a hypothesis test,
not a parameter estimate, from the beginning.  This leads naturally to
the following question.

\subsection{Which model or theory is correct?}
\label{sec:goodness}

This question is a critical one for the scientific method.
Astronomers typically do not address it quantitatively but \emph{want}
to do so.  I will separate the general question ``which model is
correct?'' into two: 1) ``does the model explain the data?'', the
\emph{goodness-of-fit} problem; and 2) ``which of two (or more) models
better explains the data?'', the \emph{model selection} problem.  Let
us begin here with Question 1 and discuss Question 2 in the next
section.

Suppose one has performed a parameter estimation and determined the
parameter region(s) containing a large fraction of the probability
density.  Before making any conclusions from the application of a
statistical model to a data set, an investigator should assess the fit
of the model to make sure that the model can explain adequately the
important aspects of the data set.  \emph{Model checking}, or
assessing the fit of a model, is a crucial part of any statistical
analysis. Serious misfit, failure of the model to explain important
aspects of the data that are of practical interest, should result in
the replacement or extension of the model. Even if a model has been
assumed to be final, it is important to assess its fit to be aware of
its limitations before making any inferences.

The posterior predictive check (PPC) is a commonly-used Bayesian model
evaluation method \citep[e.g.][Chap. 6]{Gelman.Carlin.ea:95}.  It is
simple and has a clear theoretical basis.  To apply the method, one
first defines a set of discrepancy measures. A discrepancy measure,
like a classical test statistic, measures the difference between an
aspect of the observed data set and the theoretically predicted data
set.  Let $\model$ denote the model under consideration.  Practically,
a number of predicted data sets are generated from
$P(\data|\param^\ast,\model)$ with $\param^\ast$ selected from the
posterior distribution. Any systematic differences between the
observed data set and the predicted data sets indicate a potential
failure of the model to explain the data.  For example, one may use
the distribution of a discrepancy measure based on synthetic data
generated from the posterior distribution to estimate a Bayesian
p-value for the true data under the model hypothesis.  The p-value in
this context is simply the cumulative probability for the discrepancy
statistic.  A p-value in the tails of the predicted discrepancy-measure
distribution suggests a poor fit to the data. By using a variety of
different discrepancy statistics, one's understanding of \emph{how}
the model does not fit the data is improved. See Section \ref{sec:ppc} for
more detail.

Another approach attempts to fit a non-parametric model to the data.
If the non-parametric model better explains the data than the fiducial
model, one rejects the fiducial model as a good fit.  A procedure for
assessing the model families will be described in the next section.  A
naive implementation of this idea is difficult, requiring a second
high-dimensional MCMC simulation to infer the posterior distribution
for the non-parametric model and a careful specification of the prior
distribution.  A clever scheme for doing this
\citep{VerdinelliWasserman:1998} is described in Section \ref{sec:VW}.

\subsection{Model selection and Bayes factors}
\label{sec:modelsel}

We often have doubts about our parametric models, even those that fit.
This is especially true when the models are phenomenological rather
than the results of \emph{first-principle} theories.  Therefore, we
need to estimate which competing model better represents the data.
Astronomers are becoming better versed in the more traditional
statistical \emph{rejection} tests but astronomers often really want
\emph{acceptance} tests.  Bayes factors provide this: one can
straightforwardly evaluate the evidence {\em in favour} of the null
hypothesis rather than only test evidence for rejecting it.  {\em
  Bayes factors} are the dominant method for Bayesian model selection
and are analogous to likelihood ratio tests
\citep[e.g.][]{Jeffreys:61,Gelman.Carlin.ea:95,Kass.Raftery:95}.
Rather than using the posterior extremum, one marginalises over the
parameter space to get the marginal probability of the data under each
model or hypothesis.  The ratio of the likelihood functions
marginalised over the prior distributions provides evidence in favour
of one model specification over another.  In this way, the Bayesian
approach naturally includes, requires in fact, that one's prior
knowledge of the model and its uncertainties be included in the
inference.  Although this dependence on the prior probability
sometimes criticised as a flaw in the Bayesian approach, one's prior
belief will invariably influence one's interpretation of a statistical
finding and should be carefully quantified.  The Bayesian framework
allows the scientist to describe and incorporate prior beliefs
quantitatively.  In addition, the method demands that the scientist
thoughtfully characterise prior assumptions to start.  Such discipline
will improve the quality of any scientific conclusions and provide an
explicit statement of the scientists prior assumptions for others to
examine.

Mathematically, Bayes factors follow from applying Bayes Theorem to a
space of models or hypotheses.  Let $P(\model)$ be our prior belief in
Model $\model$, and let $P(\data|\model)$ be the probability of
observing $\data$ under the assumption of Model $\model$.  The Bayes
Theorem tells us that the probability of Model $\model$ having the
observed $\data$ is
\begin{equation}
  P(\model|\data) = \frac{P(\model)P(\data|\model)}{P(\data)}
  \label{eq:bf1}
\end{equation}
where $P(\data)$ is some unknown normalisation constant. However, one
may use equation (\ref{eq:bf1}) to compute the relative probability of
two competing models, $\model_i$ and $\model_j$:
\begin{equation}
  \frac{P(\model_1|\data)}{P(\model_2|\data)}
  = \frac{P(\model_1)}{P(\model_2)}
  \frac{P(\data|\model_1)}{P(\data|\model_2)}
  \label{eq:bf2}
\end{equation}
without reference to the unknown normalisation.  The left-hand side of
equation (\ref{eq:bf2}) may be interpreted as the posterior odds ratio
of Model 1 to Model 2.  Similarly, the first term on the right-hand
side is the prior odds ratio.  The second term on the right-hand side
is called the Bayes factor.  Most often, one does not assert a
preference for either model and assigns unity to the prior odds ratio.

To define Bayes factors explicitly in terms of the posterior
distribution, suppose that one observes data $\data$; these may
comprise many observations or multiple sets of observations.  One
wishes to test two competing models (or hypotheses) $M_1$ and $M_2$,
each described by its own set of parameters, $\param_1$ and
$\param_2$.  One would like to know which of the following likelihood
specifications is better: $M_1: L_1(\data|\param_1)$ or $M_2:
L_2(\data|\param_2)$, given the prior distributions $P_1(\param_1)$
and $P_2(\param_2)$ for $\param_1$ and $\param_2$. The Bayes Factor
$B_{12}$ is given by
\begin{equation}
  B_{12} = \frac{P(\data|M_1)}{P(\data|M_2)} = \frac{\int
    \,P_1(\param_1|M_1)P_1(\data|\param_1, M_1)d\param_1}{\int
    \,P_2(\param_2|M_2)P_2(\data|\param_2, M_2)d\param_2}.
  \label{eq:BayFac}
\end{equation}
If $B_{12}>1$, the data indicate that $M_1$ is more likely than $M_2$
and vice verse.  Harold Jeffreys (1961, App.\ B\nocite{Jeffreys:61})
suggested the often-used scale for interpretation of $B_{12}$ in
half-unit steps in $\log B_{12}$ (see Table \ref{tab:jeffreys}). This
provides a simple-to-use, easily discussed criterion for the
interpretation of Bayes factors.

\begin{table}
  \caption{Jeffreys' table}
  \begin{tabular}{lll}
    $\log B_{12}$& $B_{12}$ & 	Strength of evidence \\ \hline
    $<0$ 	& $<1$ & 	Negative (supports $M_2$) \\
    0 to 1/2 	& 1 to 3.2 & 	Barely worth mentioning \\
    1/2 to 1	& 3.2 to 10 & 	Positive \\
    1 to 2	& 10 to 100 &	Strong \\
    $>2$ 	& $>100$ & 	Very strong \\ \hline
  \end{tabular}
  \label{tab:jeffreys}
\end{table}

Bayes factors are very flexible, allowing multiple hypotheses to be
compared simultaneously or sequentially. The method selects between
models based on the evidence from data without the need for
nesting\footnote{Two models are \emph{nested} if they share the same
  parameters and one of them has at least one additional parameter.}.
On the other hand, classical hypothesis testing gives one hypothesis
(or model) preferred status (the \emph{null hypothesis}) and only
considers evidence against it; the Bayes factor approach is
considerably more general.  The posterior probability for competing
models can be evaluated over an ensemble of data and used to decide
whether or not a particular family of models should be preferred.
Similarly, common parameters can be evaluated over a field of
competing models with appropriate posterior model probabilities
assigned to each.  A tutorial illustrating this can be found in the
BIE documentation.

Given all of these advantages, why are Bayes factors not more commonly
used?  There are two main difficulties. First, multidimensional
integrals are difficult to compute.  Following equation
(\ref{eq:BayFac}), one needs to evaluate an integral of the form:
$P(\data) = \int P(\param)P(\data|\param)d\param$. For a real world
model, the dimensionality of $\param$ is likely to be $>10$.  Such a
quadrature is infeasible using standard techniques.  On the other
hand, a typical MCMC calculation has generated a large number of
evaluations of the integrand at considerable expense.  Can one use the
posterior sample to evaluate the integral?

\citet{Raftery:95} suggests a \emph{Laplace-Metropolis} estimator that
uses the MCMC posterior simulation to approximate the marginal density
of the data using Laplace's approximation (see Raftery op. cit. for
details).  In practice, this is only accurate for nearly Gaussian (or
\emph{normal}) unimodal posterior distributions.  As part of the BIE
development, \citet{Weinberg:12,Weinberg.etal:2013} described two new
approaches for evaluating the marginal likelihood from the
MCMC-generated posterior sample and both of these are implemented in
the BIE (see Section \ref{sec:evidence}).  In short, the BIE together
with recent advances for computing the marginal likelihood makes the
wholesale computation of Bayes factors feasible in many cases of
interest.

A second well-known difficulty is the sensitivity of Bayes factors to
the choice of prior.  Most commonly, researchers feel that vague
priors are more appropriate than informative priors.  This leads to an
inconsistency known as the Jeffreys--Lindley paradox
\citep{Lindley:1957}, which shows that vague priors result in
overwhelming odds for or against a hypothesis by varying the parameter
that controls the vagueness (e.g. extending the range of an arbitrary
uniform distribution).  This apparent problem has led researchers to
seek Bayesian hypothesis tests that are less sensitive to their prior
distributions.  Conversely, one should not expect a vague prior to
yield a sensible model comparison.  Rather, the prior should be used
to express prior belief in a theory and, therefore, the resulting
hypothesis test should be sensitive to the prior.  In addition, the
prior specification should not be more informative than the
likelihood; this will result in strong bias. This sensitivity implies
that the theory implicit in the model is informed by one's background
knowledge.  Nonetheless, the prior knowledge is difficult to quantify
and I would still advocate testing a variety of prior distributions
consistent with one's prior knowledge.  This may be tested through
direct sensitivity analyses, such as resimulation with chains at
different resolutions and approximate priors.

Alternatively, one may \emph{condition} a vague prior using Bayesian
update for a small subset of new data or previously acquired data.
That is, the resulting posterior distribution inferred from the small
subset of data and the vague prior may be characterised and used as a
prior distribution on the remainder of the data.  This has been used
productively to classify galaxy type using the BIE.
\citet{Yoon.etal:13} show that this technique greatly improves the
reliability of Bayesian decision process.

Regardless of one's viewpoint, the BIE project currently provides a
useful platform for investigating the use of Bayesian model comparison
and hypothesis testing and, hopefully, it will help pave the way for
new applications.  In some cases, computing the Bayes factor will be
infeasible.  For these, the BIE includes an MCMC algorithm that
selects between models as part of the posterior simulation (reversible
jump) as described in Section \ref{sec:RJ}.

\subsection{Bayesian update}
\label{sec:bayes_update}

Suppose that our data consists of two components $\data = (\data_1,
\data_2)$ and $P(\param)$ is is our prior on the parameters $\param$
of our model or theory. Then:
\begin{eqnarray}
  P(\param|\data_1, \data_2) &\propto& P(\param) P(\data_1, \data_2|\param) \\
  &\propto& P(\param|\data_1) P(\data_2|\param, \data_1)
\end{eqnarray}
In other words, the following are equivalent: 1) updating our believe
in the prior $P(\param)$ by treating $(\data_1, \data_2)$ as a single
observation; and 2) updating the prior $P(\param)$ with respect to the
first observation $\data_1$ producing posterior the $P(\param|\data_1)
\propto P(\data_1|\param) P(\param)$, which serves as the prior for
the second observation $\data_2$.  This procedure is known as the
\emph{Bayesian update}. The Bayesian update is key to the solution of
the Monte Hall problem: the prior distribution is updated by the hosts
answer to your selection of the door that hopefully hides the prize.

These ideas lead to an incremental procedure based on Bayes theorem
for updating our belief in a theory or hypothesis.  Suppose that we
begin by inferring the probability of $\param$ given the first data
set $\data_1$:
\begin{equation}
  P(\param |\data_1) \propto  P(\param) P(\data_1|\param)
\end{equation}
Based on the character of $P(\param |\data_1)$, we obtain additional
data $\data_2$ to provide additional constraint.  We then, update the
Bayesian belief from $\data_2$ using prior $P (\param|\data_1 )$:
\begin{eqnarray}
  P (\param |\data_1 ,\data_2) &\propto& P (\param |\data_1) P(\data_2
  | \param ,\data_1) \nonumber \\
  &\propto& P(\param) P(\data_1|\param) P(\data_2|\param,\data_1).
\end{eqnarray}
Clearly, this procedure may be continued iteratively.  In addition,
such a situation is natural when data arriving sequentially,
i.e. $\data_1,\data_2, \ldots, \data_n$ and we wish to update or
belief or update our knowledge of an unknown parameter.  For many
problems the likelihood function does not depend on the previously
obtained data, and the last term in the equation above simplifies:
$P(\data_2|\param,\data_1) = P(\data_2|\param)$.

For complex astronomical models, it pays to have a way of reusing the
information obtained from the expensive computations necessary to
characterise the posterior distribution. The Bayesian update fits the
bill.  For example, a parameter inference for a complex model
describing a the formation of galaxies leading to the present day
population based on the distribution of galaxy luminosities (a
semi-analytic model) may constrain certain parameters in the model but
not others.  For a high-dimensional model, vast regions of parameter
space are likely to have extremely low posterior probability. Then,
the subsequent addition of different data, perhaps in a wave band
sensitive to the processes described by the unconstrained parameters
can use the first inference as prior knowledge.  This has a practical
advantage: the later inference has knowledge of what parameter values
are plausible and this expedites the sampling.  Moreover, learns
directly how the new data alters the belief in the model parameters
based on the first data set.

\subsection{Observational requirements}
\label{sec:obsreq}

The probability of the data given the parameter vector and the model,
$P(\data|\param,\model)$ or the likelihood function, is fundamental to
any inference, Bayesian or otherwise.  Meaningful inferences demand
that the data presentation include all of the information necessary
for the modeller to compute $P(\data|\param,\model)$ accurately and
precisely.  The more direct the construction of $P(\data|\param,
\model)$ from the physical theory, i.e. the less information lost in
modelling the acquired observations, the easier it is to calculate
$P(\data|\param,\model)$, leading to a higher quality result.  In
other words, the more the data is ``reduced'' through summary
statistics and ``cleaned'' by applying complex filters, the less
information remains and the greater the impact of difficult-to-model
correlations.  This is somewhat contrary to standard practice where
the presentation of scatter diagrams of summary statistics is the
norm.

In addition, astronomers often quote their error models in the form of
uncorrelated standard errors.  The customary expectation is that each
datum, typically a data bin or pixel, should be within the range
specified by the error bar most of the time.  Quoted error bars are
often inflated to make this condition obtain. This leads to a number
of fundamental flaws that makes the error model (and therefore the
data) unsuitable for Bayesian inference:
\begin{enumerate}
\item Binned and pixelated data are nearly always correlated for
  ``cleaned'' or ``reduced'' observations.  For example, a flat-field
  photometric correction and sky-brightness removal correlates the
  pixels of an image over its entire scale.  There are many additional
  sources of indirect correlations.  Parameter estimations are often
  sensitive to these correlated excursions in the data values and
  ignoring these correlations will lead to erroneous inferences.  Data
  archivists can facilitate accurate inferences by providing
  correlation matrices for all error models.
\item Selection effects must be modelled in the likelihood function
  and, therefore, these effects must be well specified by the
  archivist to facilitate straightforward computation.  For example,
  consider a multiband flux-limited source catalogue.  A
  colour-magnitude or Hess diagram in two flux bands will have a
  non-rectangular boundary owing to the flux limit.  Although this is
  a simple example, selection effects may be terribly difficult to
  model; consider spatial variations in source completeness owing to
  the diffraction spikes from bright stars.
\item Astronomers tend to use historically familiar summary data
  representations that inadvertently complicate the computation of
  $P(\data|\param,\model)$.  Continuing the previous example, the
  magnitude-magnitude diagram contains the same information as the
  colour-magnitude diagram but the selection effects lie along
  flux-level boundaries.  For a more complicated example, consider the
  Tully-Fisher diagram.  The input data set may contain flux limits,
  morphology selections, image inclination cuts, redshift range
  limits, just to name a few.
\end{enumerate}

In summary, data processing and reduction, correlates the data
representation and can hide selection effects; all of these complicate
the computation of $P(\data|\param,\model)$ and renders the modelling
process difficult and potentially unreliable.  Rather, one should
endeavour to separate each source of error, carefully specifying the
underlying acquisition process for every observational campaign.  For
example, each pixel datum in a digital image may be characterised by
the observed data number (dn), gain and bias, read noise, thermal
background fraction, etc.  Together, their combination yields an error
model.  Even if a first-principle process cannot be described, an
empirical distribution or process description will be helpful to
modellers.  For example, a distribution of deviations for measured
pixel values relative to a reference calibration field may be used as
an error model.  Ideally, the data representation should be as close
to the acquired form as possible.  In cases where the archiving or
presentation of source data is impractical, the production of a
correlation matrix is essential.

For an example, the effect of data correlation has been explored by
\citet{lu.etal:10,Lu.etal:2012}.  They describe the parameter
inference for a semi-analytic model of galaxy formation conditioned on
a galaxy mass function with both correlated and uncorrelated data
bins.  The differences in the posterior distributions for these two
cases is dramatic. When the error model is in doubt, the sensitivity
of the inference to the error model can be investigated in the
Bayesian paradigm by putting prior distributions on parameters of the
error models that describe their uncertainty, marginalising over those
hyperparameters, and comparing with the original posterior.  Although
this is more expensive and rarely done, one should consider performing
such sensitivity analyses regularly.

\section{The BIE software design and workflow}
\label{sec:design}

The BIE was designed to free the scientist from focusing on the
technical complexity of a Bayesian inference for large projects with
significant computational investment.  Most often, a scientist begins
a new inference problem with little or no experience with the features
of the posterior distribution and, the identification of most
effective sampling strategy, therefore, requires experimentation.  As
in other platforms, the BIE provides a variety of commonly used
algorithms, as well as new ones derived by us for specific research
problems.

The BIE is implemented using object-oriented patterns in C++.  See
Section \ref{sec:software} for details. This allow the mathematical
relationships between the various numerical procedures that implement
the Bayesian concepts to be reflected and maintained by the software
structure.  This provides both guidance to the user by preventing
specification errors and a natural structure for concurrent software
development.  The object-oriented approach allows the easy creation of
hybrid approaches, e.g. combining several existing algorithms, and
changing computational methodology without changing the initially
specified inference problem.  This approach naturally encourages reuse
of previous implementations and this increases the robustness and
speed of development.  For example, the implementation of the core
parallel sampling algorithms and the likelihood functions for common
data types may be computed once and for all.  Then, any new variants
and methods will be available for use by others.

All of the Bayesian tasks described in Section \ref{sec:wants_needs}
depend on the full posterior distribution, not just the peak parameter
value or summary characterisation.  Indeed, all of Bayesian analysis
is based on the \emph{algebra} of probability distributions,
e.g. equation (\ref{eq:Bayes}).  A goal of the BIE is a direct
incorporation of this algebra into the software-enabled workflow,
described below.  To this end, the BIE provides tools for efficiently
sampling and characterising features of the posterior distribution.
Our main tool for estimating the posterior density in high dimension
is a kernel density estimator based on the metric ball tree
construction.  These estimates may be used performing Bayes updates as
described in Section \ref{sec:bayes_update}.  Combined with the
persistent store for BIE inferences described in Sections
\ref{sec:design} and \ref{sec:software}, density estimations may be
saved, recalled and reused as prior distributions for new simulations
or in direct Bayesian updates (see Section \ref{sec:bayes_update}).
Finally, because methodology changes and evolves, the BIE must be
extensible without making previous results obsolete.

The BIE persists its data and internal structure using the BOOST
(\url{http://www.boost.org}) serialisation library.  In essence, this
allows the running or stopped simulation to be written to disk and
read from disk at the byte level.  This allows, for example, the
kernel density estimates resulting from an expensive characterisation
of a posterior distribution to be saved and used any number of times
for, e.g., later Bayesian updates or to inform new inferences.
Furthermore, the BOOST serialisation library automatically records
version numbers which allows the code to evolve but maintain backward
compatibility with older cached sessions.

Finally, the BIE is designed with flexible data handling including
consumer-producer data streams with matched likelihood functions.
These are easily extended to include new data types and allow compound
likelihood function specifications for multiple data types.  The goal
is a reuse of previous specifications in new ways for addressing new
scientific problems.

In summary, the software design was motivated by four main desires: 1)
to provide a computational platform optimised for massively parallel
computing clusters; 2) to provide an extensible platform that
encourages experimentation with the latest inferential mathematical
and computational tools without requiring ground-up implementation; 3)
to provide a high-level interface for defining an inference using the
algebra of probability distributions that efficiently uses prior
results; and 4) to provide a way of storing or \emph{persisting} the
details or state of the computational inference so that these may be
reused later and archived to preserve one's computational investment.

\begin{figure*}
  \centering
  \includegraphics[height=0.9\textheight]{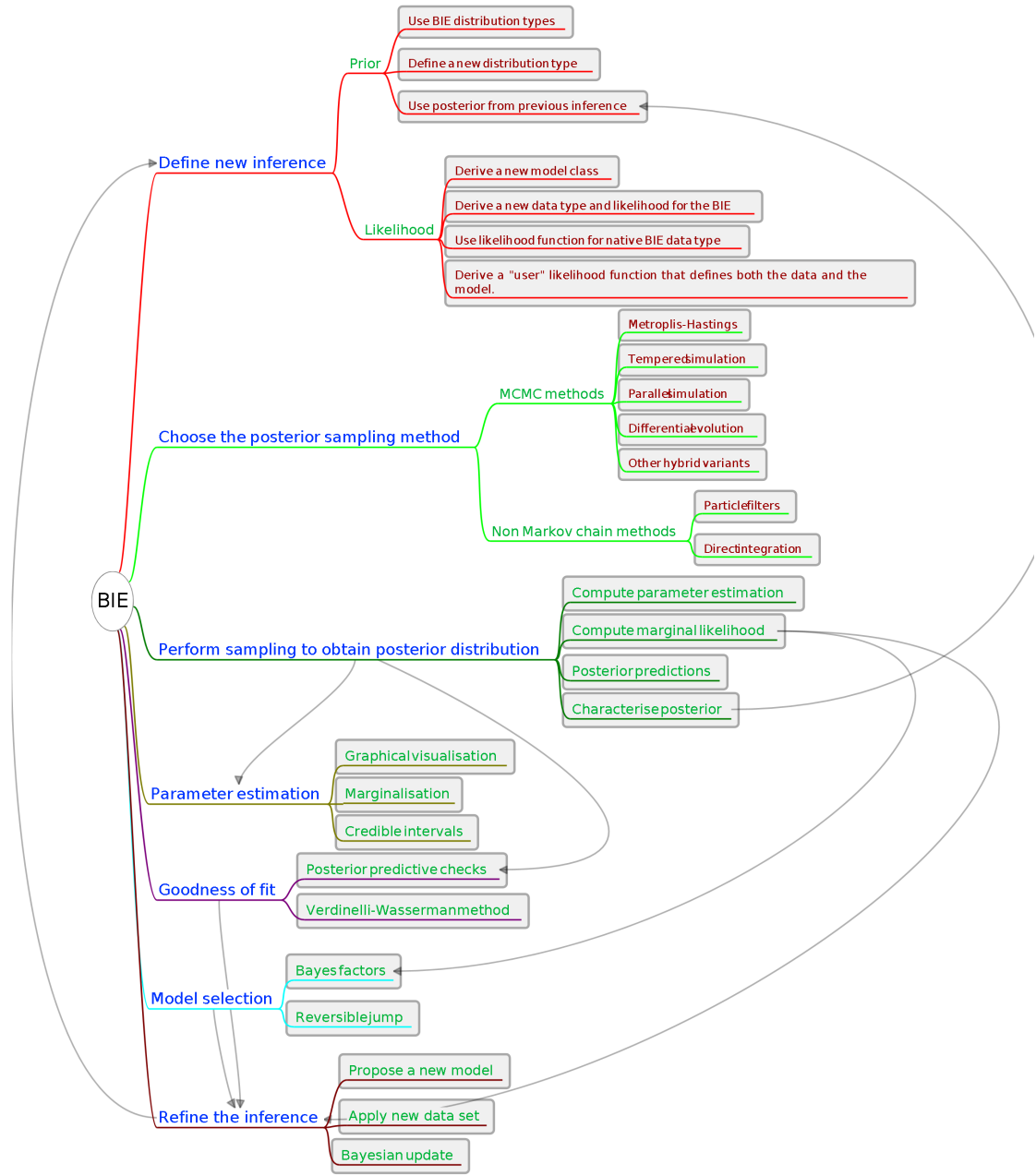}
  \caption{BIE workflow and collaboration diagram.  The solid coloured
    paths indicate parts of the inference (blue) and choices and
    options or each part (green and maroon) for each part of the
    inference.  The grey curved arrows show implicit connections
    between the parts within the BIE that may be made asynchronously
    as part of the scientific investigation.}
  \label{fig:workflow}
\end{figure*}

The overall BIE-enabled workflow is diagrammed in Figure
\ref{fig:workflow}.  The main tasks for the Bayesian inference are
listed in blue.  Each child node (green) lists the sub tasks (green).
Each of these are defined within the BIE as classes which may be
invoked by calling the BIE library or using the command-line parser.
These sub tasks may have a number of possible options (maroon, not all
of which are shown here).  For example, a variety of MCMC sampling
algorithms are available in the BIE; once the prior distribution and
likelihood function of the data at hand are specified, the quality and
features of the sampled posterior distribution may be compared within
the BIE with no additional effort on scientist's part.  The posterior
may then be used in parameter estimation, goodness-of-fit analyses,
for model section and for Bayesian update (as indicated by the grey
curves).  All of these refine the import of the inference and provide
new avenues for further observations and hypothesis testing,
reflecting the scientific method.

\section{Case studies}
\label{sec:examples}

\subsection{Semi-analytic galaxy formation models: BIE-SAM}
\label{sec:bs}

Many of the physical processes parametrised in semi-analytic models of
galaxy formation remain poorly understood and under specified.  This
has two critically important consequences for inferring constraints on
the physical parameters: 1) prior assumptions about the size of the
domain and the shape of the parameter distribution will strongly
affect any resulting inference; and 2) a very large parameter space
must be fully explored to obtain an accurate inference.  Both of these
issues are naturally tackled with a Bayesian approach that allows one
to constrain the theory with data in a probabilistically rigorous way.
In addition, for many processes in galaxy formation, competing models
have been proposed but not quantitatively compared.  Bayes-factor
analyses enable the probabilistic assessment of competing models based
on their ability to explain the same data.  In \citet{lu.etal:10}, we
presented a semi-analytic model (SAM) of galaxy formation in the
framework of Bayesian inference and illustrated its performance on a
test problem using the BIE; we call the combined approach BIE-SAM.
Our sixteen-parameter semi-analytic model incorporates all of the most
commonly used parametrisations of important physical processes from
existing SAMs including star formation, SN feedback, galaxy mergers,
and AGN feedback.

\begin{figure*}
  \centering
  \includegraphics[width=0.75\textwidth]{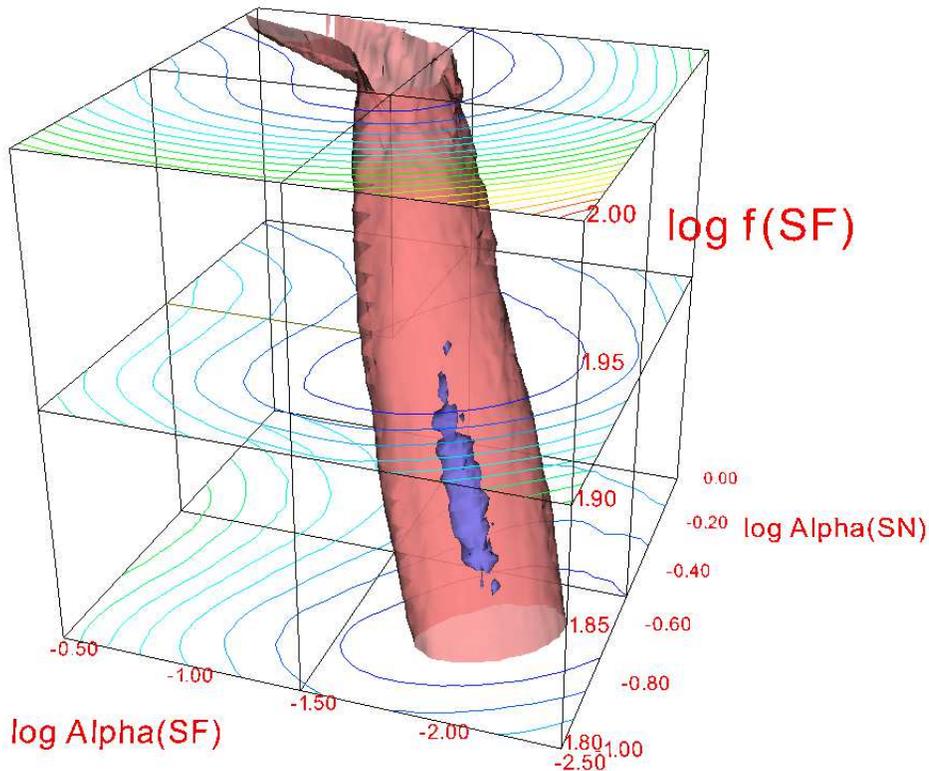}
  \caption{The likelihood function for 3 out of the 13 free parameters
    in the BIE-SAM from \citet{lu.etal:10}: the star-formation
    threshold surface density $f_{SF}$, the star-formation efficiency
    power-law index $\alpha_{SF}$, and the supernova feedback energy
    fraction $\alpha_{SN}$.  The blue (red) surfaces enclose
    approximately 10\% (67\%) of the density.  See \citet{lu.etal:10}
    for additional details.}
  \label{fig:likevol}
\end{figure*}

To demonstrate the power of this approach, the thirteen of these
parameters that can be constrained by the K-band luminosity function
were investigated in \citet{Lu.etal:2012}.  We find that the posterior
distribution has a very complex structure and topology, indicating
that finding the best fit by tweaking model parameters is improbable.
As an example, Figure \ref{fig:likevol} describes isosurfaces of the
posterior distribution in three of thirteen dimensions.  The surfaces
have a complex geometry and are strongly inhomogeneous in any
parameter direction.  Moreover, the posterior clearly shows that many
model parameters are strongly covariant and, therefore, the inferred
value of a particular parameter can be significantly affected by the
priors used for the other parameters.  As a consequence, one \emph{may
  not} tune a small subset of model parameters while keeping other
parameters fixed and expect a valid result.

Apropos the discussion in Section \ref{sec:obsreq}, by using synthetic data
to mimic systematic uncertainties in the reduced data, we also have
shown that the resulting model parameter inferences can be
significantly affected by the use of an incorrect error model.  We
used a synthetically-generated binned stellar mass function and
performed two inferences: one with a realistic covariance and one with
no off-diagonal covariance.  The contours with the full covariance
matrix are more compact, but there are also noticeable changes in the
shape and orientation of the posterior distribution.  This clearly
demonstrates that an accurate analysis of errors, both sampling errors
and systematic uncertainties, are crucial for observational data, and
conversely, a data-model comparison without an accurate error model is
likely to be erroneous.

The method developed here can be straightforwardly applied to other
data sets and to multiple data sets simultaneously.  In addition, the
Bayesian approach explicitly builds on previous results by
incorporating the constraints from previous inferences into new data
sets; the BIE is designed to do this automatically.  

\subsection{Galphat}
\label{sec:galphat}

\begin{figure*}
\begin{minipage}{\textwidth}
  \includegraphics[width=\textwidth]{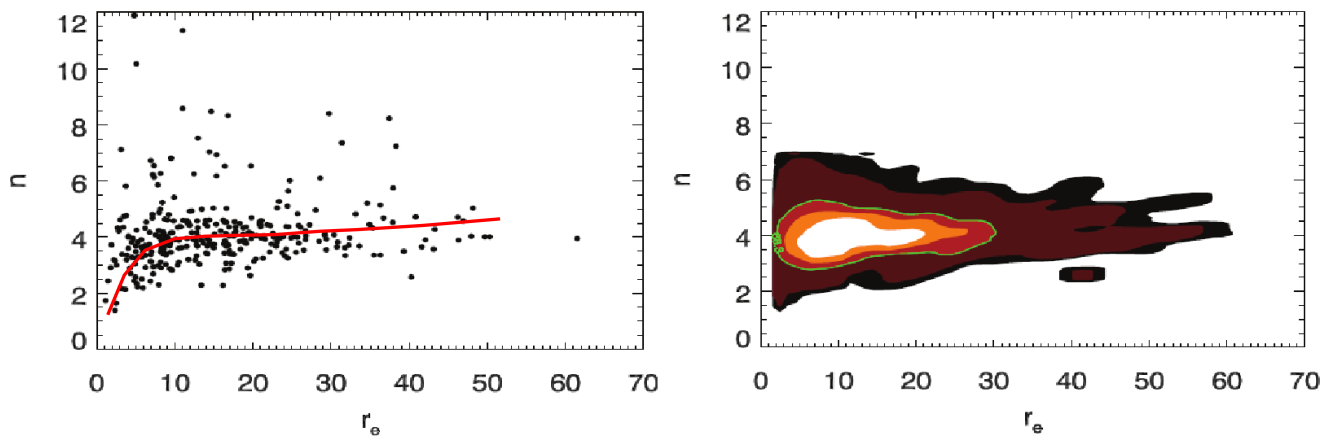}
  \caption{The size--S\'ersic index relation inferred from a
    synthetically-generated sample of elliptical galaxy images Left: a
    scatter plot using the best-fit parameters.  The red curve shows a
    smooth fit to the ridge-line of the points. Right: the marginal
    posterior density for the same parameters.  While the left-hand
    plot suggests that small galaxies have low concentrations, the
    right-hand plot of posterior density correctly reveals that this
    trend is an artefact of the model-fitting procedure.  These
    figures originally appeared in \citet{Yoon.etal:11}.}
  \label{fig:galphat1}
\end{minipage}
\end{figure*}

\begin{figure*}
\centering
\begin{minipage}{0.85\textwidth}
  \includegraphics[width=\textwidth]{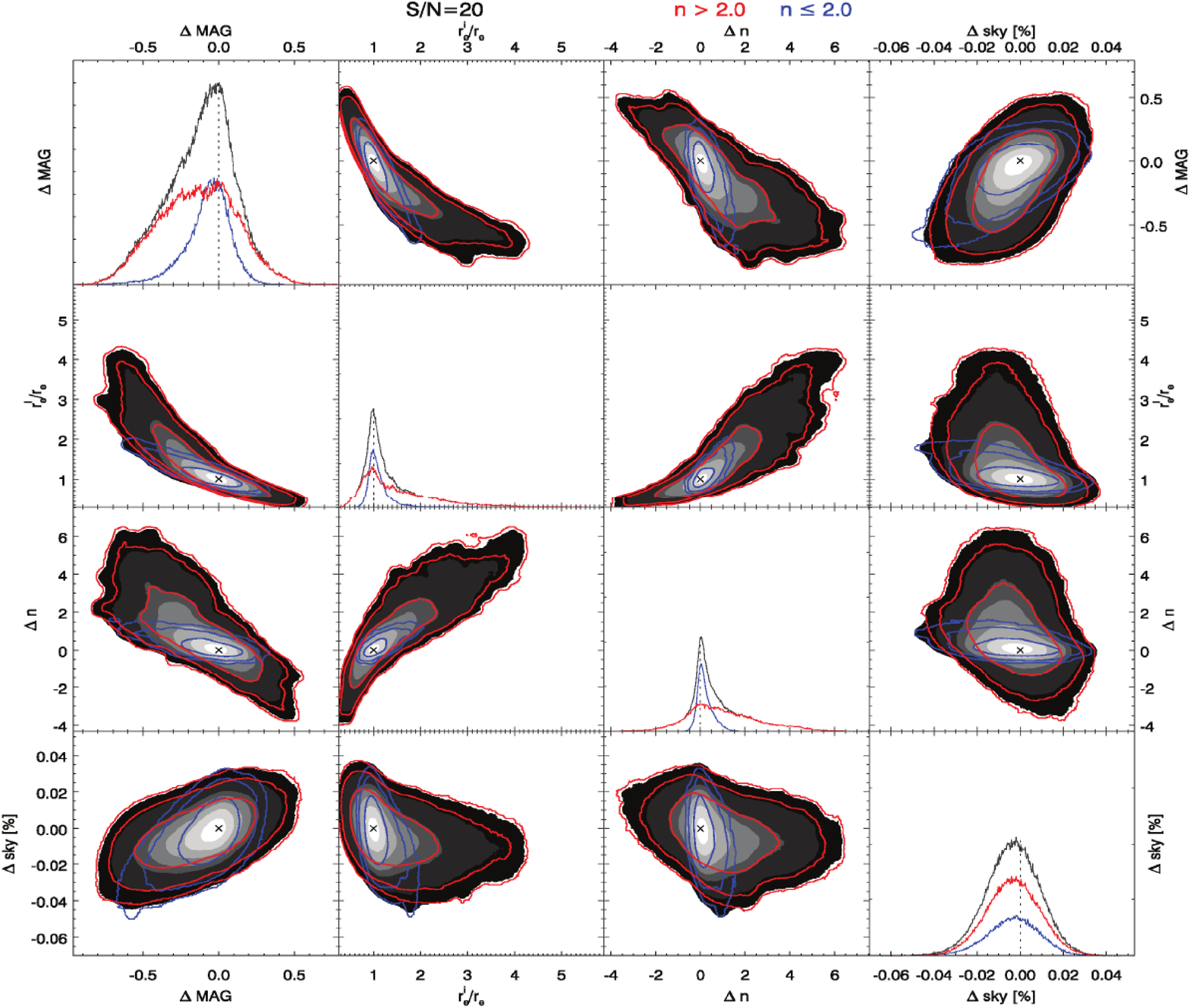}
  \caption{Marginal posterior densities for 100 galaxies with a
    signal-to-noise ratio of 20 and randomly sampled S\'ersic model
    parameters.  The values are magnitude difference from the input
    values ($\Delta$MAG), galaxy half-light radius scaled by the input
    values ($r^i_e/r_e$), the S\'ersic index difference ($\Delta n$)
    and the sky value difference ($\Delta$sky in percent).  The blue,
    red and grey curves and contours show galaxies with $n\ge2$, $n<2$
    and the total sample, respectively.  The parameter covariance
    depends on both the signal-to-noise ratio and the S\'ersic index.}
  \label{fig:galphat2}
\end{minipage}
\end{figure*}

\citet{Yoon.etal:11} describes Galphat (GALaxy PHotometric
ATtributes), a Bayesian galaxy image analysis package built for the
BIE, designed to efficiently and reliably generate the posterior
probability distribution of model parameters given an image.  From the
BIE point of view, both Galphat and the BIE-SAM are likelihood
functions with internally defined data.  A general binned data type
(images and histograms) will be native in the next BIE release. The
Galphat likelihood function is designed to produce high-accuracy
photometric predictions for galaxy models in an arbitrary spatial
orientation for a given point-spread function (PSF).  Accurate
predictions in both the core and wings of the image are essential for
reliable inferences.  The pixel predictions are computed using an
adaptive quadrature algorithm to achieve a predefined error tolerance.
The rotation and PSF convolution are performed on sub-sampled grids
using an FFT algorithm.  Galphat can incorporate any desired galaxy
image family. For speed, we precompute subsampled grids for each model
indexed by parameters that influence their shape.  The desired
amplitude and linear scale are easily computed through coordinate
transformations. To enable this, the images are stored as
two-dimensional cumulative distributions.  Our current implementation
uses \citet{Sersic:1963} models for disk, bulge and spheroid
components, however, this approach is directly applicable to any model
family.

Using the various tempering algorithms available in the BIE, our tests
have demonstrated that we can achieve a steady-state distribution and
that the simulated posterior will include any multiple modes
consistent with the prior distribution.  Given the posterior
distribution, we may then consistently estimate the credible regions
for the model parameters.  We show that the surface-brightness model
will often have correlated parameters and, therefore, any hypothesis
testing that uses the ensemble of posterior information will be
affected by these correlations. The full posterior distributions from
Galphat identify these correlations and incorporate them in subsequent
inferences.  Our work to date has extensively explored two models: a
one-S\'ersic component model with eight parameters and a two-S\'ersic
component model with twelve parameters.

These issues are illustrated in Figures \ref{fig:galphat1} and
\ref{fig:galphat2}.  Figure \ref{fig:galphat1} shows the
size--S\'ersic index relation inferred from a synthetically-generated
sample of elliptical galaxy images.  The left-hand panel shows the
traditional scatter diagram of maximum posterior parameter values;
that is, this figure plots the effective radius and S\'ersic index for
the maximum likelihood value obtained for each image fit.  The red
curve is a smooth estimate of the trend.  The right-hand panel shows
the inferred distribution based on the full posterior distribution of
the ensemble after marginalising over all of the parameters but the
effective radius and the S\'ersic index.  The left-hand panel
\emph{incorrectly} suggests that smaller galaxies are less
concentrated while the right-hand panel \emph{correctly} reveals that
the size and concentration are uncorrelated.  Figure
\ref{fig:galphat2} illustrates the correlation between parameters for
low-concentration galaxies (S\'ersic index $n\le2$) in blue and
high-concentration galaxies ($n>2$) in red with the total in grey
scale.

We can use posterior simulations over ensembles of images to test, for
example, the significance of cluster and field environments on
galaxies as evidenced in their photometric parameters, such as the
correlation between bulge-to-disk ratio and environment.  A more
elaborate example might include models for higher angular harmonics of
the light distribution and we could determine the support for these in
the data using Bayes factors (Section \ref{sec:evidence}).  Recent
work successfully demonstrates the use of Bayes factors with the BIE
algorithms described in Section \ref{sec:evidence} to classification
of galaxy images \citep{Yoon.etal:13}.

\subsection{Performance}
\label{sec:perf}

Both examples used the \emph{differential evolution} algorithm (see
Section \ref{sec:DE}), chosen after testing straight
Metropolis-Hastings (Section \ref{sec:met_hast}), tempered chains
(Section \ref{sec:sim_temp}), and parallel chains (Section
\ref{sec:PT}).  As previously emphasised, the choice of algorithm is
problem dependent.  However, one of the key frustrations of applying
Metropolis-Hastings MCMC algorithms to real-world high-dimensional
posterior distributions is the choice of a function form for
transition probability that facilitates exploring the posterior space.
Most often, the shape of the posterior and even the range of the
meaningful parameters values are not known.  This requires a number of
inefficient tuning runs whose posterior distributions may be
characterised to set an efficient transition probability function.  

On the other hand, the differential evolution algorithm automatically
supplies a tuned proposal distribution from its own ensemble of
chains.  The downside of differential evolution is that exploration of
the posterior space may be poor.  To this end, we have augmented the
standard differential evolution algorithm by simulated tempering as
described in Section \ref{sec:DE}.  However, tempering is expensive.
Each of typically 20 temperature levels uses 10 iterations per
level. We typically take 10 to 20 standard steps between tempering
steps.  This implies that the tempered differential evolution requires
10 times more evaluations than the straight differential evolution
algorithm!  Once the chain is converged, however, each state of all
the converged chains are good posterior samples.  However expensive,
this approach does work when others do not for complex posterior
distributions such as that described in Section \ref{sec:bs}).

For Galphat, the posterior distribution is sufficiently regular that
straight differential evolution is sufficient.  We typically run 16
chains for each galaxy until 200,000 converged posterior samples are
obtained.  The convergence is diagnosed using the Gelman-Rubin $R$
statistic \citep{Gelman.Rubin:92} which compares the sample variance in the
chain and among the chains.  In many runs, a few individual chains in
isolated local maximum become stuck; that is, if local maximum
contains a single chain, the ensemble of chains may not be capable of
readily forming a combination of difference vectors that presents an
improved proposal (see eq. \ref{eq:de}). The number of such outliers
is typically 1 out of 16 and not more than 4.  The BIE uses a modified
Grubbs statistic \citep{Grubbs:1969} to automatically detect such
outliers and eliminate them from the final posterior sample.  The
auto-correlation length for the generated Markov chain is
approximately 30 and reaches statistical equilibrium in approximately
1,000 iterations.

We have also explored a hierarchical Bayesian update algorithm for
improving convergence.  We iteratively add image information using a
hierarchy of successively aggregated images.  Beginning with the most
aggregated image (Level 0) one computes the posterior,
$P(\param|\data_0)$.  The posterior for the next level (Level 1) is
the $P(\param|\data_1) \propto P(\param|\data_0)
[P(\data_1|\param)/P(\data_0|\param)]$ and so on.  This reduces the
time by a factor of two to four depending on the level of aggregation
by accelerating convergence.  For example, 20,000 converged MCMC
sample is obtained within 2 hours and 40 minutes for non-hierarchical
and hierarchical data structure, using 8 Opteron 1533MHz cores.

For the semi-analytic model inferences, we use our
\emph{tempered}-differential evolution algorithm (to run the MCMC
simulation with 128 chains in parallel. The initial states of the
chains are randomly distributed in parameter space according to the
prior probability distribution. We terminate the simulation after
16,000 iterations, when a sufficiently large number of states are
collected to summarise the marginalised posterior.  The
auto-correlation length for this simulation is of order the tempering
interval, typically 30.  Again, the Gelman-Rubin statistic monitors
the convergence of the MCMC simulation.  For a particular but typical
test example, we find that 123 chains that are well mixed; the other 5
chains are outliers.  The individual chains are initialised from the
prior distributions and are all widely dispersed at the beginning of
the simulation. The mixed chains gradually converge to a high
probability mode after about 3,000 iterations.  In contrast, the
outlier chains do not converge, but wander around in low probability
regions. The simulations are kept running for 16,000 iterations, even
though most of the chains have ``burned-in'' after 4,000
iterations. We take the consecutive 12,000 steps of the 123 converged
chains, about 1.5 million states, to summarise the marginalised
posterior probability distributions of the model parameters.

\section{Discussion and summary}
\label{sec:summary}

Advances in digital data acquisition along with storage and retrieval
technologies have revolutionised astronomy.  The promise of combining
these vast archives have led to organisation frameworks such as the
National Virtual Observatory (NVO) and the International Virtual
Observatory Alliance (IVOA).  To realise the promise of this vast
distributed repository, the modern astronomer needs and wants to
combine multi-sensor data from various surveys to help constrain the
complex processes that govern the Universe.  The necessary tools are
still lacking, and the Bayesian Inference Engine (BIE) was designed as
a research tool to fill this gap.  This paper outlines the motivation,
goals, architecture, and use of the BIE and reports our experience in
applying MCMC methods to observational and theoretical Bayesian
inference problems in astronomy.

Most researchers are well-versed in the identifying best parameters
for a particular model for some data using the maximum likelihood
method.  For example, consider the fit of a surface brightness model
to galaxy images.  Parameters from the maximum-likelihood solutions
are typically plotted in a scatter diagram and correlations between
these parameters are interpreted physically.  However, plotting
scatter diagrams from multiple data sources inadvertently mixes error
models and selection effects.  Section \ref{sec:paramest} described
the pitfalls of this approach. Rather, the astronomer wants to test
the hypothesis that the data is correlated with a coefficient larger
than some predetermined value $\alpha$, a complex \emph{hypothesis
  test}.  However, without incorporating the correlations imposed by
both the theoretical model, the error model, and the selection
process, the significance of the test is uncertain.  Similarly, the
astronomer needs methods of assessing whether a posited model is
correct.  In Section \ref{sec:wants_needs}, I divided these needs into
two categories: goodness-of-fit tests (Section \ref{sec:goodness}) and
model selection (Section \ref{sec:modelsel}).  As an example of the
former, the astronomer may have found the best parameters using
maximum likelihood, but does the model fully explain all of the
features in the data?  If it does not, one must either modify or
reject the model before moving on to the next step.  As an example of
the latter, suppose an inference results in two parameter regions or
multiple models that explain the data.  Which model \emph{best}
explains the data?

All of these wants and needs---combining data from multiple sources,
estimating the probability of model parameters, assessing goodness of
fit, and selecting between competing models---are naturally addressed
in a single probabilistic framework embodies in \emph{Bayesian
  inference}. In particular, Bayesian inference provides a data-first
discipline that demands that the error model and selection effects are
specified by the probability distribution for the data given the model
$\model$, $P(\data|\param,\model)$, colloquially known as the
likelihood function $L(\data|\param,\model)$.  Prior results including
quantified expert opinion are specified in the prior probability
function $P(\theta|\model)$.  The inferential computation may be
incremental: the data may be added in steps and new or additional
observations may be motivated at each step, true to the scientific
method.  In the end, this approach may be generalised to finding the
most likely models in the generalised space of models; this leads to
goodness-of-fit and model comparison tests.

For scientists, the ideal statistical inference is one that lets the
data ``speak for themselves.''  This is achievable in some cases.  For
example, estimating a small number of parameters given a large data
set tends to be independent of prior assumptions.  On the other hand,
hypothesis tests of two complex models may depend sensitively on prior
assumptions.  For a trivial example, some choices of parameters may be
unphysical even though they yield good fits and should be excluded
from consideration.  Moreover, if two competing models fit the data
equally well, any hypothesis test will be dominated by prior
information.  Inferences based on realistic models of astronomical
systems will often lie between these two extremes.  For these cases, I
advocate Bayesian methods because they precisely quantify both the
scientists' prior knowledge and the information gained through
observation.  In other words, we allow the data to ``speak for
themselves'' but in a ``dialect'' of our choosing.  Philosophy
aside, Bayes theorem simply embodies the law of conditional
probability and provides a rigorous framework for combining the prior
and derived information.

With these advantages comes a major disadvantage: Bayesian inference
is computationally expensive!  An inference may require a huge sample
from the posterior distribution and real-world computation of
$P(\data|\param,\model)$ is often costly.  Moreover, naive MCMC
algorithms used for sampling the posterior distribution converge
unacceptably slowly for distributions with multiple modes, and
advanced techniques to improve the mixing between modes are needed,
increasing the expense.  Finally, Bayesian inference requires
integrals over typically high-dimensional parameter spaces.  For
example, Bayes factors (Section \ref{sec:modelsel}) require the computation
of the marginal likelihood:
\[
P(\data|\model) = \int d\mathbf{\theta}\, P(\param|\model)
P(\data|\param,\model).
\]
Evaluation of this integral suffers from the curse of dimensionality.

Bayesian inference is, essentially, an algebra of probability
distributions for describing one's evolving beliefs in the light of
new data, models, and hypotheses.  Effective use of Bayesian inference
by scientists requires high-performance tools that manipulate
probability distributions directly, allowing the scientist to focus on
their meaning.  The elegance and promise of Bayesian inference
motivated us to attempt a computational solution and this became the
BIE project.  The algorithms and techniques described here, all and
more available in the BIE, have proved useful to address the
complications found in research problems. In short, the BIE fills a
gap between tools developed for small-scale problems or those designed
to test new algorithms and a computational platform designed for
production-scale inference problems typical of present-day
astronomical survey science.  Its primary product is a representation
of the posterior distribution to be used for parameter estimation and
model selection.  Other Bayesian applications, such as non-parametric
inference and clustering, should be possible with little modification,
but have not yet been investigated.  The BIE is designed to run on
high-performance computing clusters, although it will also run on
workstations and laptops.

Several new algorithms have been developed for and currently appear
only in the BIE.  Based on the features of complex posterior
distributions from inferences for high dimensional phenomenological
models, we implemented a self-tuning Metropolis-Hastings algorithm
based on a genetic algorithm known as differential evolution
\citep{Storn.Price:97, Storn:99, TerBraak:06}.  To this, we added
tempering following \citet[see Section \ref{sec:sim_temp} for
additional details]{Neal:96}.  This enhanced parallel algorithm
dramatically improves the accurate sampling of multimodal
high-dimensional posteriors.  In addition, a fair and computationally
efficient representation of the posterior distribution is at the heart
of the BIE's strategy to embody the algebra of probability
distributions.  We do this in one of three ways.  First, a crude by
often used tool is the multivariate normal based on the covariance
matrix in parameter space; in our experience, this is almost always
too inaccurate to be of value for predictions, but it does provide a
quick characterisation.  Second, we provide a kd-tree-based density
estimator that computes the density from the volume local to the
evaluation point; this approach is computational efficient and
provides a non-parametric estimate but the number of posterior
evaluations required for an accurate density grows exponentially in
the dimensionality.  Thirdly, our most accurate approach uses a kernel
density estimator based on a metric ball tree; we currently use a
Euclidean metric whose coefficients are scaled inversely to the
variance in each dimension.  For $N$ points, the construction
complexity is $N(\log_2N)^2$ and the evaluation complexity is $\log_2
N$ and overhead to find the nearest neighbours in the surrounding
volume.

The open object-oriented architecture allows for cross-fertilisation
between researchers and groups with both mathematical and scientific
interests, e.g. both those developing new algorithms and those
developing new astronomical models for different applications.  New
classes contributed by one become available to all users after an
upgrade.  Approaches implemented by the one user's new classes may
solve an unanticipated set of problems for other users.  In this way,
the BIE is a distributed collaborative system similar to packages like
IDL or modules in Python. I anticipate that users with a variety of
technical skill levels will use the BIE.  By reusing and modifying
supplied examples, a user's model of a likelihood function can be
straightforwardly added to the system without any detailed knowledge
of the internals.  A user's new model becomes an internally-documented
first-class object within the BIE by following the examples as
templates.  There is also room for the experienced programmer to
improve the low-level parallelism or implement more efficient
heuristics for likelihood evaluations.  The BIE includes a full
persistence subsystem to save the state and data for running a MCMC
simulation.  This facilitates both checkpointing and recovery as well
as later use of inferred posterior distributions in new and unforeseen
ways.  A future version will implement a built-in database for
warehousing results including the origin and history of both the data
and computation, along with labels, notes and comments.  Altogether,
this will constitute an electronic notebook for Bayesian inference.

The BIE provides the astronomer an organisational and computational
schema that discriminates between models or hypotheses and suggests
the best use of scarce observational resources.  In short, if we can
make better use of the interdependency of our observations given our
hypotheses, then we can generate a far clearer picture of the
underlying physical mechanisms.  In many ways, the Bayesian approach
emulates the empirical process: begin with a scientific belief or
expectation, add the observed data and then modify the extent of that
belief to generate the next expectation. In effect, as more and more
data is added to the model, the accurate predictions become reinforced
and the inaccurate ones rejected.  This approach relieves the
scientist from directly confronting the complex interdependencies
within the data since those interdependencies are automatically
incorporated into the model.  Although our scientific motivations are
astronomical, the BIE can be applied to many different complex systems
and may even find applications in areas as diverse as biological
systems, climate change, and finance.

%%%%%%%%%%%%%%%%%%%%%%%%%%%%%%%%%%%%%%%%%%%%%%%%%%%%%%%%%%%%%%%%%%%%%%%%%%
% Acknowledgments
%%%%%%%%%%%%%%%%%%%%%%%%%%%%%%%%%%%%%%%%%%%%%%%%%%%%%%%%%%%%%%%%%%%%%%%%%%
\section{Acknowledgements}

This material is based upon work supported by the National Science
Foundation under Grant Nos. 0611948 and 1109354 and by NASA AISR
Program through award NNG06GF25G.  The initial development of the BIE
was previously funded by NASA's Applied Information and System
Research (AISR) Program.  I thank Neal Katz and Eliot Moss for
comments on an early draft of this manuscript Alison Crocker, Neal
Katz, Yu Lu, and Michael Petersen for comments on the final draft.

%%%%%%%%%%%%%%%%%%%%%%%%%%%%%%%%%%%%%%%%%%%%%%%%%%%%%%%%%%%%%%%%%%%%%%%%%%
% Bibliography
%%%%%%%%%%%%%%%%%%%%%%%%%%%%%%%%%%%%%%%%%%%%%%%%%%%%%%%%%%%%%%%%%%%%%%%%%%
\label{sec:ref}

\appendix

\section{Solutions provided by the BIE}
\label{sec:bie}

\subsection{Computing the posterior distribution}

\label{sec:comp_post}

This section presents four MCMC sampling algorithms of increasing
complexity included in the BIE.  All of these have been heavily
tested.  I begin with a description and some motivation for the
standard Metropolis-Hastings algorithm.  This simple easy-to-use and
implement algorithm often fails to converge and is difficult to tune
for complicated high-dimension distributions.  The next three sections
introduce modifications that circumvent these pitfalls.

The original BIE emphasised Markov chain Monte Carlo (MCMC) methods.
MCMC excels because it circumvents the exponential $n^d$ scaling of
traditional exhaustive sampling, where $n$ is the number of knots per
dimension and $d$ is the number of dimensions.  Conversely, the
inherent difficulty in assessing the convergence of the Markov chain
has led some to revisit and improve direct methods.  The current
version includes, in addition, particle filters and direct quadrature
methods.  A later version will include the nested sampler
\citep{Skilling:06} and its variants \citep{Feroz.Hobson:08}.

\subsubsection{The Metropolis-Hastings algorithm}
\label{sec:met_hast}

Metropolis-Hastings is the most well-known of MCMC algorithms
\citep{Metropolis.Rosenbluth.ea:53,Hastings:70}.  This algorithm
constructs a Markov chain that generates states from the
$P(\param|\data,\model)$ distribution after a sufficient number of
iterations.  Its success requires that the Markov chain satisfy both
an ergodicity and a detailed balance condition.  The ergodicity
condition ensures that at most one asymptotic distribution exists.
Ergodicity would fail, for example, if the chain could cycle back to
its original state after a finite number iterations.  Ergodicity also
requires that all states with positive probability be visited
infinitely often in infinite time (called \emph{positive recurrence}).
For general continuous state spaces, ergodicity is readily achieved.
The detailed balance condition ensures that the chain admits at least
one asymptotic distribution.  Just as in kinetic theory or radiative
transfer, one defines a transition process that takes an initial state
to a final state with some probability.  If the Markov chain samples
the desired target distribution $P(\param)=P(\param|\data,\model)$,
detailed balance demands that the rate of transitions
$\param\rightarrow\param^\prime$ is the same as the rate of
transitions from $\param^\prime\rightarrow\param$.

One may state this algorithm explicitly as follows.  Let $P(\param)$
be the desired distribution to be sampled and
$q(\param,\param^\prime)$ be a known, easy-to-compute transition
probability between two states.  Given $\param$, the distribution
$q(\param,\param^\prime)$ is a probability distribution for
$\param^\prime$.  Let $a(\param,\param^\prime)$ be the probability of
accepting state $\param^\prime$ given the current state $\param$. In
short, one can show that if the detailed balance condition
\begin{equation}
P(\param)q(\param,\param^\prime)a(\param,\param^\prime) =
P(\param^\prime)q(\param^\prime,\param)a(\param^\prime,\param)
\label{eq:detailed_bal}
\end{equation}
holds, then the Markov chain will sample $P(\param)$.  It is
straightforward to verify by substitution that acceptance probability
\begin{equation}
a(\param,\param^\prime) = \min\left\{1,
  [P(\param^\prime)q(\param^\prime,\param) /
  P(\param)q(\param,\param^\prime)]\right\}
\label{eq:MH}
\end{equation}
solves this equation \citep[for additional discussion see][]{liu:04}.
Equation (\ref{eq:detailed_bal}) has the same form as well-known
kinetic rate equations as follows.  Given the probability of a
transition over some time interval from a state A to some other state
B of a physical system and the corresponding reverse reaction, then
the equilibrium condition for $N=N_A + N_B$ systems distributed in the
two states is:
\[
N_A P(A\rightarrow B) = N_B P(B\rightarrow A).
\]
In equation (\ref{eq:detailed_bal}), the probability densities
$P(\param)$ and $P(\param^\prime)$ play the r\^ole of the occupation
numbers $N_A$ and $N_B$ and the product of the transition and
acceptance probabilities play the r\^ole of the probabilities
$P(A\rightarrow B)$ and $P(B\rightarrow A)$.

The transition probability $q(\param,\param^\prime)$ is often chosen
to facilitate the generation of $\param^\prime$ from
$\param$. \citet{Metropolis.Rosenbluth.ea:53} introduced a kernel-like
transition probability $q(\param,\param^\prime) = {\bar q}(\param
- \param^\prime)$ where ${\bar q}(\cdot)$ is a density.  This has the
easy-to-use property of generating $\param^\prime = \param +
\mathbf{\xi}$ where $\xi\sim {\bar q}$. (The notation $\theta\sim
P(\theta)$ expresses that the distribution function of the variate
$\theta$ is $P(\theta)$.)  Further, if ${\bar q}$ is symmetric,
i.e. ${\bar q}(\mathbf{z}) = {\bar q}(-\mathbf{z})$, then equation
(\ref{eq:MH}) takes the simple form
\begin{equation}
a(\param,\param^\prime) = \min\left\{1,
  \frac{P(\param^\prime)}{P(\param)}\right\}
\label{eq:MH2}
\end{equation}
The BIE provides two symmetric distributions for ${\bar q}$: a
multivariate normal and a uniform or \emph{top-hat} distribution.  The
user may easily add new distributions as appropriate.  Each element of
$\param$ is scaled by a supplied vector of \emph{widths}, ${\bf w}$.
The choice of ${\bf w}$ is critical to the performance of the
algorithm. If the width elements are too large,
$P(\param^\prime)/P(\param)$ will tend to be very small and proposed
states will rarely be accepted.  Conversely, if the width elements are
too small, the new state frequently will be accepted and successive
states will be strongly correlated.  Either extreme leads to process
that is slow to reach equilibrium.  The optimal choice is somewhere in
between the two.  As the dimensionality of the parameter space grows,
specifying the optimal vector of widths a priori is quite difficult.
I will address this difficulty in Section \ref{sec:DE}.

\subsubsection{Tempered transitions}
\label{sec:sim_temp}

In addition to the inherent difficulties associated with \emph{tuning}
the transition probability, the Metropolis-Hastings state can easily
be trapped in isolated modes, between which the Markov chain moves
only rarely.  This prevents the system from achieving detailed balance
and, thereby, prevents sampling from the desired target distribution
$P(\theta)$.  There are a number of techniques for mitigating this
so-called \emph{mixing problem}.  For the BIE, I adopted a synthesis
of Metropolis-coupled Markov chains \citep{Geyer:91} and a simulated
tempering method proposed by \citet{Neal:96} called \emph{tempered
  transitions}.  To sample from a distribution $P_0(\param)\equiv
P(\param)$ with isolated modes, one defines a series of $n$ other
distributions, $P_1(\param),\ldots, P_n(\param)$, with $P_k$ being
easier to sample than $P_{k-1}$.  For example, one may choose
\begin{equation}
P_k(\param) \propto P_0^{\beta_k}(\param)
\label{eq:pup}
\end{equation}
with $1=\beta_0>\beta_1>\cdots>\beta_{n-1}>\beta_n>0$.  This
construction has a natural thermodynamic interpretation. One may write
$P_0^{\beta_k}(\param) = e^{\beta_k\log(P_0)} \equiv
e^{\log(P_0)/T_k}.$ The distribution with temperature $T_0=T=1$ is the
original distribution. \emph{Hotter} distributions have higher
temperature $T_k>T_0$ and are over-dispersed compared with the
original \emph{cold} distribution. In other words, taking a
distribution function to a small fractional power decreases the
dynamic range of its extrema.  In the limit $T_k\rightarrow\infty$,
$P_k$ becomes uniform.  Next, the method defines a pair of base
transitions for each $k$, $\tup_k$ and $\tdown_k$, which both have
$P_k$ as an invariant distribution and satisfy the following mutual
reversibility condition for all $\param$ and $\param^\prime$:
$P_k(\param) \tup_k(\param, \param^\prime) =
\tdown_k(\param^\prime, \param) P_k (\param^\prime)$.

A tempered transition first finds a candidate state by applying the
base transitions in the sequence $\tup_1\cdots\tup_n$.  After each
upward transition, new states are sampled from a broader distribution.
In most cases, this liberates the candidate state from confinement by
the mode of the initial state.  This is then followed by a series of
downward transitions $\tdown_n\cdots\tdown_1$.  This candidate state
is then accepted or rejected based on ratios of probabilities
involving intermediate states.  Thermodynamically, each level $k$
corresponds to an equilibrium distribution at temperature $T_k$.
Therefore, the upward or downward transitions correspond to
\emph{heating} or \emph{cooling} the system, respectively. One chooses
the maximum temperature to be sufficiently high to \emph{melt} any
structure in the original \emph{cold} posterior distribution that
would inhibit mixing.  Since this value depends on the features of
$P(\theta)$ that are not known a priori, I choose $T_n$ by checking
the distribution of $\theta$ for $P_n$ with various values of $T_n$.
Our tests have shown that this algorithm works remarkably well for a
variety of different inference problems.

Explicitly, the algorithm proceeds as follows.  The chain begins in state
${\hat\param}_0$ and obtains the candidate state, ${\check\param}_0$,
as follows.  For $j=0$ to $n-1$, generate ${\hat\param}_{j+1}$ from
${\hat\param}_{j}$ using ${\hat T}_{j+1}$.  Set
${\check\param}_n={\hat\param}_n$.  Then, for $j=n$ to $1$, generate
${\check\param}_{j-1}$ from ${\check\param}_{j}$ using ${\check
  T}_{j}$.  The candidate state, ${\check\param}_0$, is then accepted
with probability
\begin{equation}\small
  a \equiv \min\left[1, \frac{P_1({\hat\param}_0)}{P_0({\hat\param}_0)} \cdots
    \frac{P_n({\hat\param}_{n-1})}{P_{n-1}({\hat\param}_{n-1})} \cdot
    \frac{P_{n-1}({\check\param}_{n-1})}{P_n({\check\param}_{n-1})} \cdots
    \frac{P_0({\check\param}_0)}{P_1({\check\param}_0)} \right].
  \label{eq:alphaST}
\end{equation}
Although equation (\ref{eq:alphaST}) looks complicated, the derivation
in \citet{Neal:96} shows that it immediately follows by recursively
applying equations (\ref{eq:detailed_bal}) and (\ref{eq:MH}).  Then,
owing to the mutual reversibility condition, the $\tup$ and $\tdown$
dependence in equation (\ref{eq:alphaST}) cancels.  If the candidate
state is not accepted, the next state of the Markov chain is the same
as the original state, ${\hat\param}_0$.  In practice, I `burn-in' the
chain at each level $j$ for $M$ iterations, with $M=20$ typically.
Note that each $P_i$ occurs an equal number of times in the numerator
and denominator in equation (\ref{eq:alphaST}).  Therefore, the
acceptance probability can be computed without knowledge of the
normalisation constants for these distributions.  If the acceptance
probability is to be reasonably high, properly-spaced intermediate
distributions will have to be provided that gradually interpolate from
$P_0$ to $P_n$. Thermodynamically, this corresponds to an adiabatic
increase of the heat-bath temperature followed by an adiabatic return
to the original temperature.  

Many readers will be familiar with the idea of \emph{simulated
  annealing} \citep{Kirkpatrick.etal:83}.  Each step of a simulated
annealing algorithm proposes to replace the current state by a state
randomly constructed from the current state by a transition
probability.  The magnitude of the excursion allowed by the transition
probability is controlled by a temperature-like parameter that is
gradually decreased as the simulation proceeds.  This prevents the
state from being stuck at local maxima.  The tempered transitions
algorithm is heuristically similar to simulated annealing with the
important additional property of obeying detailed balance.

\subsubsection{Parallel tempering}
\label{sec:PT}

The parallel tempering algorithm inverts the order of the previous
algorithm: it simultaneously simulates $n$ chains, each with target
distribution $P_j$ (eq. \ref{eq:pup}) and proposes to swap states
between adjacent members of the sequence at predefined intervals.  The
high temperature chains are generally able to sample large volumes of
parameter space, whereas low temperature chains, may become trapped in
local probability maxima.  Parallel tempering achieves good sampling
by allowing the systems at different temperatures to exchange states
at very different locations in parameter space.  The higher
temperature chains often achieve detailed balance quickly and
accelerate the convergence of the lower temperature chains.  Thus,
this method may allow a simulation to achieve detailed balance even in
the presence of widely separated modes.  In some situations, parallel
tempering outperforms tempered transitions with a lower overall
computational cost.  In addition, the parallel tempering algorithm is
trivially parallelised by assigning each chain its own process.  The
tempered transitions algorithm is intrinsically serial and parallel
efficiency is only obtained if the likelihood computation is
parallelisable.

The parallel tempering algorithm proceeds as follows.  At each step, a
pair of adjacent simulations in the series is chosen at random and a
proposal is made to swap their parameter states.  The new state is
accepted using the following Metropolis-Hastings criterion.  Let the
$j^{th}$ iterate of the state in the $k^{th}$ chain be denoted as
$\param^{[k]}_j$.  The swap is accepted with probability
\begin{equation}
  a = \min\left[1,
    \frac{P_k(\param^{[k+1]}_j|\data,\model)P_{k+1}(\param^{[k]}_j|\data,\model)}
    {P_k(\param^{[k]}_j|\data,\model)P_{k+1}(\param^{[k+1]}_j|\data,\model)}
  \right],
\end{equation}
where $P_k(\param|\data,\model)$ is the posterior probability of
$\param$ given the data $\data$ for chain $k$ and model assumptions
$\model$.  Final results are based on samples from the $\beta_0 = 1$
chain. As in the tempered transitions algorithm, the high-temperature
states will mix between separated modes more efficiently, and
subsequent swapping with lower-temperature chains will promote their
mixing.

The analogue thermodynamic system here is an \emph{array} of systems
with the same internal dynamics at different temperatures.  At higher
temperatures, strongly `forbidden' states are likely to remain
forbidden but valleys between multiple modes are likely to be more
easily crossed.  In contrast to tempered transitions
(Section \ref{sec:sim_temp}), the proposed transitions in parallel tempering
are \emph{sudden} exchanges of state between systems of possibly
greatly different temperature.  As with tempered transitions, the
algorithm obeys the detailed balance equation.

\subsubsection{Differential evolution}
\label{sec:DE}

\newcommand{\paramj}{\boldsymbol{\theta}^{[j]}}

Real-world high-dimensional likelihood functions often have complex
topologies with strong anisotropies about their maxima (see Section
\ref{sec:bs}, Fig. \ref{fig:likevol}).  Difficulties in tuning the
Metropolis-Hastings transition probability to achieve both a good
acceptance rate and good mixing plagues high-dimensional MCMC
simulations of the posterior probability.  This problem affects all of
algorithms discussed up to this point.  \citet{TerBraak:06} introduced
an MCMC variant of a genetic algorithm called \emph{differential
  evolution} \citep{Price.Storn:97, Storn.Price:97, Storn:99}.  This
version of differential evolution uses an ensemble of chains, run in
parallel, to adaptively compute the Metropolis-Hastings transition
probability.  In all that follows, I will refer to the MCMC variant as
simply differential evolution.

The algorithm proceeds as follows.  Assume that our ensemble has $n$
chains to start, e.g., initialised from the prior probability
distribution.  Each chain has the same target distribution
$P(\param)$.  The original differential evolution algorithm
\citep{Price.Storn:97} proposes to update member $i$ as follows:
$\paramj_p = \paramj_{R0} + \gamma(\paramj_{R1} - \paramj_{R2})$ where
$R0, R1, R2$ are randomly selected without replacement from the set
$\{1, 2, \ldots, n\}$.  The proposal vector replaces the chosen one if
$P(\paramj_p)>P(\paramj_i)$.  \citet{TerBraak:06} shows that with
minor modifications the transition probability and the acceptance
condition for differential evolution obeys detailed balance.  The new
MCMC version of the differential evolution algorithm takes the form
\begin{equation}
  \paramj_p = \paramj_i + \gamma(\paramj_{R1} - \paramj_{R2}) +
  \bf{\epsilon}
  \label{eq:de}
\end{equation}
where ${\bf\epsilon}$ is drawn from a symmetric distribution with a
small variance compared to that of the target, but with unbounded
support such as a $d$ dimensional normal distribution with very small
variance $\sigma^2$: ${\bf\epsilon} \sim {\cal N}^d(0, \sigma^2)$.
The random variate $\bf{\epsilon}$ is demanded by the recurrence
condition: the domain for non-zero values of the posterior $P$ must be
reached infinitely often for an infinite length chain.  The proposal
$\paramj_p$ is accepted as the next state for Chain $i$ with
probability
\[
a = \min\left[1,
  \frac{P({\bf\theta}_p)}{P({\bf\theta}_i)}\right].
\]
In essence, differential evolution uses the variance between a
population of chains whose distributions are converging to the target
distribution to automatically tune the proposal widths.  Although the
transition probability distribution $q(\param,\param^\prime)$ does not
have an analytic form in this application, the differential evolution
algorithm enforces symmetry through the random choice of indices, and
the distribution $q(\param,\param^\prime)$ clearly exists.

This algorithm as stated above does not address the mixing problem.
\citet{TerBraak:06} suggests including simulated tempering or
simulated annealing in differential evolution.  Along these lines, BIE
also includes a hybridised differential evolution which periodically
performs a tempered transition step for all $n$ chains in parallel.
This provides an ensemble at each temperature for the upward and
downward transitions.  As in tempered transitions, we evolve each
chain for $M$ steps at each temperature level.  A typical number of
temperature levels is fifteen, and therefore, the addition of tempering
may slow the algorithm by an order of magnitude.  Although
\emph{tempered} differential evolution is dramatically slower than the
simple differential evolution, we have found it essential for
achieving a converged posterior sample for many of our real-world
astronomical inference problems.  This method was applied to the
Bayesian semi-analytic models described in \citet{lu.etal:10}.

\subsubsection{Summary: choice of a MCMC algorithm}

I advocate performing a suite of preliminary simulations to explore
the features of one's posterior distribution with various algorithms.
My experience suggests that there is no single \emph{best} MCMC
algorithm for all applications.  Rather, each choice represents a set
of trade offs: more elaborate algorithms with multiple chains,
augmented spaces, etc., are more expensive to run but may be the only
solution for a complex posterior distribution.  Conversely, an
elaborate algorithm would be wasteful for simulating a simple
posterior distribution.  Moreover, combinations of MCMC algorithms in
multiple-chain schemes are often useful.  Finally, as we will describe
below in Section \ref{sec:evidence}, the potentially expensive
likelihood evaluations that are not accepted during the course of the
MCMC algorithm may be cached for use in performing the computational
quadrature of the marginal likelihood integral.

For distributions with complex topologies, differential evolution
relieves the scientist of the task of hand selecting a transition
probability by trial and error.  This method has the advantage of
added efficiency: states from all chains in a converged simulation
provide valid posterior samples.  However, this strategy may backfire
if the posterior is strongly multimodal because differential evolution
requires multiple chains in each mode to enable mixing between modes.
Any single chain in a discrete mode will remain forever.  Parallel
chains and similar algorithms do not have this problem.  Although
high-temperature chains from tempered methods do not sample the
posterior distribution, they do provide useful information for
importance sampling.  \citet{Yoon.etal:11} has productively used
high-temperature samples for Monte Carlo integration.

\subsection{Computation of Bayes factors and marginal likelihoods}
\label{sec:evidence}

As described in Section \ref{sec:modelsel}, the marginal likelihood
plays a key role in Bayesian model selection.  There are several
common strategies for computing the marginal likelihood.  The simplest
is direct quadrature using multidimensional cubature algorithms
\citep[e.g.][]{Berntsen.etal:1991}.  Computational complexity limits
its application to approximately four or fewer dimensions.  Secondly,
one may approximate the integrand around each well-separated mode as a
multivariate normal distribution and integrate the resulting
approximation analytically.  This is the \emph{Laplace approximation}.
It suits simple unimodal densities approximately Gaussian shape, but
the posterior distributions for many real-world problems are far from
Gaussian.  Finally, one may rewrite Bayes theorem as an expression
that evaluates the normalisation constant from the posterior sample as
the harmonic mean of the likelihood function, as will be shown below.
In short, none of the three suffices in general: direct quadrature is
most often computationally infeasible, the Laplace approximation works
well only for simple posterior distributions and the harmonic mean
estimator often has enormous variance owing to its inverse weighting
by the likelihood value \citep[see][]{Kass.Raftery:95}.  To help
address this lack, \citet{Weinberg:12} presents two
computationally-modest families of quadrature algorithms that use the
full sample posterior but without the instability of the harmonic mean
approximation \citep{Newton.Raftery:94} or the specificity of the
Laplace approximation \citep{Lewis.Raftery:97}.

The first algorithm begins with the normalised Bayes theorem:
\begin{equation}
Z\times P(\param|\data) = P(\param) P(\data|\param)
\label{eq:Pdef0}
\end{equation}
where
\begin{equation}
  Z \equiv \int_\Omega d\param\, P(\param|\data) = \int_\Omega
  d\param\,P(\param) P(\data|\param)
\label{eq:Zdef0}
\end{equation}
normalises $P(\param|\data)$ (as in eq. \ref{eq:Bayes}). The quantity
$Z$ is called the \emph{normalisation constant} or \emph{marginal
  likelihood} depending on the context.  Dividing by $P(\data|\param)$
and integrating over $\param$ we have
\begin{equation}
  Z\times \int_\Omega d\param\,\frac{P(\param|\data)}{P(\data|\param)}
  = \int_\Omega d\param\,P(\param).
\label{eq:Zdef}
\end{equation}
Since the Markov-chain samples the posterior, $P(\param|\data)$, the
computation of the integral on the left from the chain appears as an
inverse weighting with respect to the likelihood.  This is poorly
conditioned owing to the inevitable small values of $P(\data|\param)$.  However, if the integrals in equation
(\ref{eq:Zdef}) are dominated by the domain sampled by the chain, the
integrals can be approximated by quadrature over a truncated domain,
$\Omega_s$ that eliminates the small number of the chain states with
low $P(\data|\param)$.  More precisely, the integral on the
left-hand-side may be cast in the following form:
\begin{equation}
  \int_{\Omega_s} d\param\,\frac{P(\param|\data)}{P(\data|\param)} =
  \int dY M(Y).
\label{eq:Zdef1}
\end{equation}
This integral will be a good approximation to the original if the
measure function defined by
\begin{equation}
  M(Y) = \int_{1/P(\data|\param)>Y}d\param\, P(\param|\data).
    \label{eq:measure}
\end{equation}
decreases faster than $P(\data|\param)$ as
$P(\data|\param)\rightarrow0$.  Otherwise, the integral in equation
(\ref{eq:Zdef1}) does not exist and the first algorithm cannot be
used; see \citet{Weinberg:12} for details.  Intuitively, one may
interpret this construction as follows: divide up the parameter space
$\param\in\Omega_s$ into volume elements sufficiently small that
$P(\param|\data)$ is approximately constant within each volume
element.  Then, sort these volume elements by their value of
$Y(\data|\param) \equiv L^{-1}(\data|\param)$.  The probability
element $dM\equiv M(Y+dY) - M(Y)$ is the prior probability of the
volume between $Y$ and $Y + dY$.  However, if the truncated volume
forms the bulk of the contribution to equation (\ref{eq:Zdef1}), the
evaluation will be inaccurate.

To evaluate the r.h.s. of equation (\ref{eq:Zdef}), one may use the
sampled posterior distribution itself to tessellate the sampled volume
in $\Omega_s\subset\Omega$.  This may be done straightforwardly using
a space-partitioning structure.  A binary space partition (BSP) tree,
which divides a region of parameter space into two exclusive sub
regions at each node, is particularly efficient.  The most easily
implemented tree of this type for arbitrary dimension is the kd-tree
(short for k-dimensional tree).  The kd-tree algorithms split
$\mathbb{R}^k$ on planes perpendicular to one of the coordinate system
axes.  The implementation provided for the BIE uses the median value
along one of axes (a \emph{balanced} kd-tree).  I have also
implemented a hyper-octree.  The hyper-octree generalises the octree
by splitting each n-dimensional parent node into $2^n$ hypercubic
children.  Unlike the kd-tree, the hyper-octree does not split on
point location and the size of the cells is not strictly coupled to
the number of points in the sample.  In addition, the cells in the
kd-tree might have extreme axis ratios but the cells in the
hyper-octree are hypercubic.  This helps provide a better
representation of the volume containing sample points.  See
\citet{Weinberg:12} for additional details, tests, and discussion.
Approximate tessellations also may be useful.  For example, the
nearest neighbour to every point in a sample could be used to
circumscribe each point by a sphere of maximum volume such that all
spheres are non-overlapping.  Comparisons and performance details will
be reported in a future contribution \citep{Weinberg.etal:13}.

For cases where the integral in (\ref{eq:Zdef1}) does not exist or the
first algorithm provides is a poor approximation, $Z$ may be evaluated
directly using the second computational approach.  Begin by
integrating equation (\ref{eq:Pdef0}) over $\Omega_s\subset\Omega$:
\begin{equation}
  Z \times \int_{\Omega_s} d\param\, P(\param|\data) = \int_{\Omega_s}
  d\param\,P(\param) P(\data|\param).
\end{equation}
The Monte Carlo evaluation of the integral on the left-hand side is
simply the fraction of sampled states in $\Omega_s$ relative to the
entire sample: $F_{\Omega_s}\equiv\sum_{\theta_i\in\Omega_s}
1/\sum_{\theta_i\in\Omega} 1$.  The integral on the right-hand side
may be evaluated using the space-partitioning procedure described
above.  Altogether, then, one has
\begin{equation}
  Z = F_{\Omega_s}^{-1}\int_{\Omega_s} d\param\,P(\param) P(\data|\param)
  \label{eq:Zdef2}
\end{equation}
where $\Omega_s$ is ideally chosen to avoid regions of very low
posterior probability.  This method has no problems of existence for
proper probability densities.

There are several sources of error in this space partition.  For a
finite sample, the variance in the tessellated parameter-space volume
will increase with increasing volume and decreasing posterior
probability.  This variance may be estimated by bootstrap.  As usual,
there is a variance--bias trade-off in choosing the resolution of the
tiling: the bias of the probability value estimate increases and the
variance decreases as the number of sample points per volume element
increases.  Some practical examples suggest that the resulting
estimates are not strongly sensitive to the number of points per cell.

Finally, there are no in-principle restrictions on the choice of
$\Omega_s$.  Therefore, we may choose $\Omega_s$ to minimise the
variance of computing $Z$ from equation (\ref{eq:Zdef2}) by
simultaneously minimising the variance of each term.  The variance of
the first term is that of counting and the Markov chain.  The variance
of the second term depends on the shape of the integrand; it will
smallest for approximately constant regions in posterior probability,
near a peak in the posterior density.  Suppose that our target
variance is $2\epsilon^2$.  This suggest choosing $\Omega_s$ centred
at a peak in the posterior distribution with a number of points $N_s$
such that $\sqrt{\lambda/N_s}\approx\epsilon$ where $\lambda$ is the
autocorrelation length of the chain.  The second-term integral may be
evaluated by Monte Carlo or cubature and will converge quickly,
achieving a variance of $\epsilon^2$, if $P(\param)P(\data|\param)$ is
slowly varying for $\param\in\Omega_s$.  This procedure is easy to
perform and does not require any difficult numerical analysis.
Details have been described in \citet{Weinberg.etal:2013}.  We find
that the required sample size from the posterior distribution is
typically reduced by an order of magnitude over those in
\citet{Weinberg:12} using this method.

In summary, the choice between the various algorithms depends on the
problem at hand.  The Laplace approximation may be a good choice for
posterior distributions that are unimodal with light tails but this is
often not the case for real-world problems.  I investigate the
performance of the algorithms in \citet{Weinberg:12} for
high-dimensional distributions in \citet{Weinberg.etal:13}.  To date,
I have reliably evaluated $Z$ for $n\le14$ using a MCMC-generated
samples of approximately $10^6$ points with auto-correlation lengths
of approximately 20.  All of these methods are included within the BIE
currently.

\subsection{Dimension switching algorithms}
\label{sec:RJ}

When generating large sample sizes that are necessary for an accurate
computation of the marginal likelihood is impractical, one may propose
a number of different models and choose between them as part of the
Monte Carlo sampling process.  This is done by adding a discrete
indicator variable to the state to designate the active model.  The
resulting state space consists of a discrete range for the indicator
and of continuous ranges for each of the parameters in each model.
\citet{Green:95} showed that the detailed balance equation can be
formulated in such a general state space.  This allows one to propose
models of different dimensionality and thereby incorporate model
selection into the probabilistic simulation itself.  The algorithm
requires a transition probability to and from each subspace
\citep{Green:95}.

For example, suppose one has an image of a galaxy field that one would
like to model with some unknown number of distinct galaxies
$k\in[1,n]$.  The extended sample space, then, consists of $n$
subspaces, each one of which contains the parameter vectors for each
of the $k$ galaxies for each subspace.  Suppose that the current state
is in the $k=3$ subspace; that is, the current model has three
galaxies.  With some predefined probability at each step, $p(3,4)$,
the algorithm proposes a transition to $k=4$ galaxies by splitting one
of three galaxies into two separate but possibly blended components.
Similarly, with some predefined probability at each step, $p(4,3)$,
one defines the reverse transition by combining two adjacent
components into one component.  Finally, no subspace transition is
proposed with the probability $p(3,3) = 1-p(3,4)-p(4,3)$.  For model
comparison, an estimate of the marginal probability for each model $k$
follows directly from the occupation frequency in each subspace of the
extended state space.

To make this explicit following \citet{Green:95}, one first defines
reversible transitions between models in different subspaces, say $i$
and $j$.  This is accomplished by proposing a bijective function
$g_{ij}$ that transforms the parameters between subspaces
$g_{ij}(\theta^{[i]}, \phi^{[i]} ) = (\theta^{[j]} , \phi^{[j]})$, and
enforces the dimensional matching condition $d(\theta^{[i]}) +
d(\phi^{[i]}) = d(\theta^{[j]}) + d(\phi^{[j]})$ where the operator
$d(\cdot)$ returns the rank of the vector argument.  The parameter
vector $\phi^{[\cdot]}$ is a random quantity used in proposing changes
in the components and for choosing additional components when going to
a higher dimension.  The rank of $\phi^{[\cdot]}$ may be zero.  For
example, if $d(\theta^{[i]})=2$ and $d(\theta^{[j]})=1$ then one may
define $d(\phi^{[i]})=0$ and $d(\phi^{[j]})=1$.  In other words, for
the purposes of inter-dimensional transitions, each subspace is
augmented by random variates with the constraint that the
dimensionality of the augmented spaces match.

Then, if $q_{ij}(\theta^{[i]}, \phi^{[i]})$ is the probability density for
the proposed transition and $p(i, j)$ is the probability to move from
subspace $i$ to subspace $j$, the acceptance probability may be
written as
\begin{eqnarray}
&&  \alpha_{ij}(\theta^{[i]}, \theta^{[j]}) =  \\ \nonumber 
&&  \min\left\{
  1, \frac{P_j(\theta^{[j]}|\data)p(j,i)q_{ji}(\theta^{[j]},\phi^{[j]})}{
    P_i(\theta^{[i]}|\data)p(i,j)q_{ij}(\theta^{[i]},\phi^{[i]})}
  \left|\frac{\partial(\theta^{[j]},\phi^{[j]})}%
    {\partial(\theta^{[i]},\phi^{[i]})}\right| \right\}
\end{eqnarray}
where the final term in the second argument of $\min\{\cdot\}$ is the
Jacobian of the mapping between the augmented spaces, and
$P_j(\theta^{[j]}|\data)$ is the posterior probability density for
the model in subspace $j$.  The probability densities $p(i, j)$ and
$q_{ij}$ are selected based on prior knowledge of the problem and to
optimise the overall rate of convergence.  The algorithm can be
summarised as follows.  Assume that the state at iteration $i$ is in
subspace $n_i$ with parameter vector $\theta_i^{[n_i]}$. One proposes a
new state as follows:
\begin{enumerate}
\item Choose a new model $j$ by drawing it from distribution
  $p(n_i,\cdot)$. Propose a value for the parameter $\theta^{[j]}$ by
  sampling $\phi^{[n_i]}$ from the distribution
  $q_{n_ij}(\theta_i^{[n_i]} , \phi^{[n_i]})$.
\item Accept the move with probability $\alpha_{n_ij}(\theta^{[n_i]},
  \theta^{[j]})$.  
\item If the move is accepted, let $n_{i+1} = j$ and
  $\theta_{i+1}^{[n_{i+1}]} = \theta^{[j]}$.
\item If the move is not accepted, stay in the current subspace :
  $n_{i+1} = n_i$ and $\theta_{i+1}^{[n_{i+1}]} = \theta_i^{[n_i]}$.
\end{enumerate}
\citet{Green:95} named this algorithm \emph{Reversible Jump Markov
  chain Monte Carlo} (RJMCMC).  One often chooses to interleave RJMCMC
steps with some number of standard Metropolis-Hastings steps to
improve the mixing in each subspace.

Within the constraints, the transitions are only limited by one's
inventiveness.  However, I have found that the most successful
transitions are intuitively motivated by features of the scientific
problem.  Although RJMCMC enables model selection between arbitrary
families of models with various dimensionality, the convergence of the
MCMC simulation depends on the specification of the transition
probabilities and a careful tuning of the MCMC algorithm.  In
addition, I have not tried RJMCMC with multiple-chain algorithms that
propose transitions between chains.  This appears to be formally sound
but the requirement that the transitioning chains be in the same
subspace may yield very long mixing times.  Similarly, the
differential evolution algorithm (see Section \ref{sec:DE}) would require
that each subspace of interest be populated by some number of chains
all times.

\subsubsection{Example: a simple transition probability}

Consider two models, Model 1 with two real parameters: $\param^{[2]}:
(\theta_1, \theta_2) \in{\cal R}^2$ and Model 2 with one real
parameter $\param^{[1]}: \theta\in{\cal R}$.  Let us assume that the
prior probability of transition between the models is
$p(1,2)=p(2,1)=1/2$ and adopt the transformation between the
subspaces: $g_{12}(\theta_1, \theta_2) = \left[(\theta_1+\theta_2)/2,
  (\theta_1-\theta_2)/2\right] = (\theta, \phi)$.  The variable $\phi$
is distributed as $q(\phi)$.  The inverse transformation is:
$g_{21}(\theta, u) = (\theta + \phi, \theta - \phi)$.  Therefore,
given $\theta$, one draws $\phi$ from $q(\phi)$ and immediately obtain
$(\theta_1, \theta_2)$.  The acceptance probability for $(\theta_1,
\theta_2)\rightarrow\theta$ becomes
\begin{eqnarray}
  \alpha_{21} &=&
  \frac{P_1(\theta)q(\phi)}{P_2(\theta_1,\theta_2)}
  \left|\frac{\partial(\theta,
      \phi)}{\partial(\theta_1,\theta_2)}\right|
  \nonumber \\
  &=&
  {P_1\left(\frac{\theta_1+\theta_2}{2}\right)
    q\left(\frac{\theta_1-\theta_2}{2}\right)}{2P_2\left(\theta_1,
      \theta_2\right)}.
\end{eqnarray}
and for $\theta\rightarrow(\theta_1, \theta_2)$:
\begin{equation}
  \alpha_{12} =
  \frac{2P_2(\theta + \phi, \theta - \phi)}{P_1(\theta)q(\phi)}.
\end{equation}

\subsubsection{Example: mixture modelling in the BIE}

\begin{figure*}
  \centering
  \mbox{\subfigure[Model 1]{\includegraphics[width=0.25\textwidth]{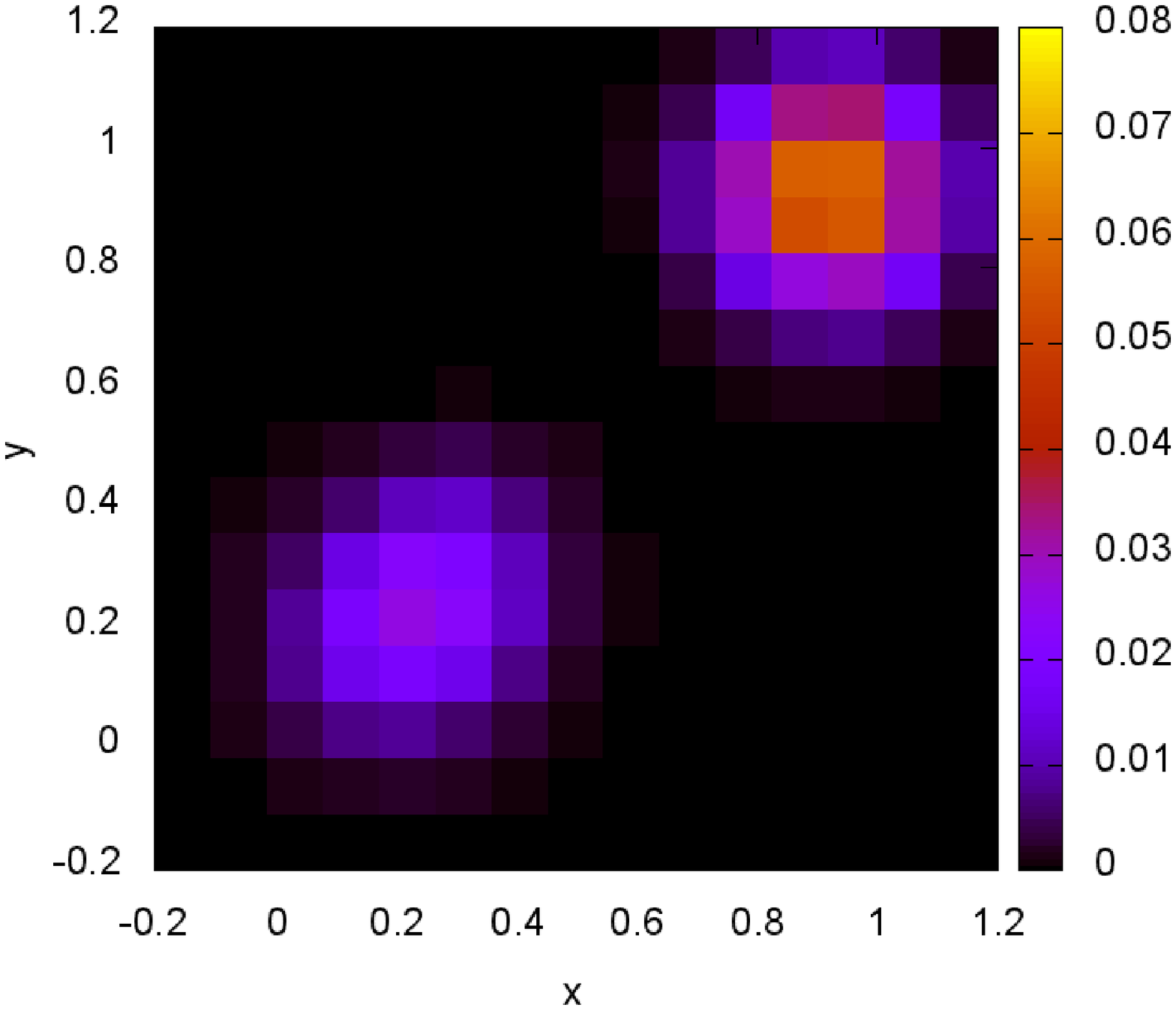}}
    \subfigure[Model 2]{\includegraphics[width=0.25\textwidth]{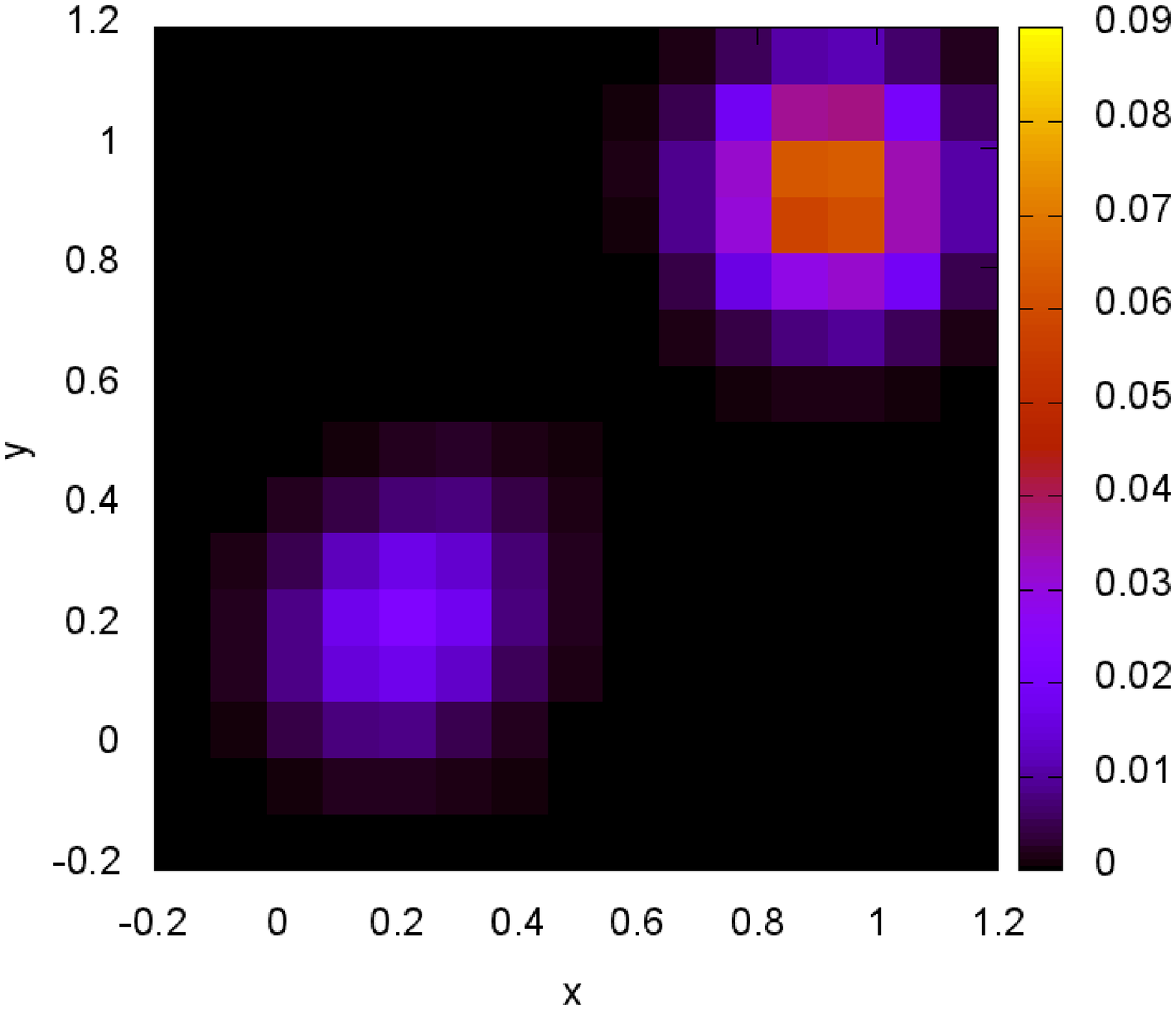}}
    \subfigure[Model 3]{\includegraphics[width=0.25\textwidth]{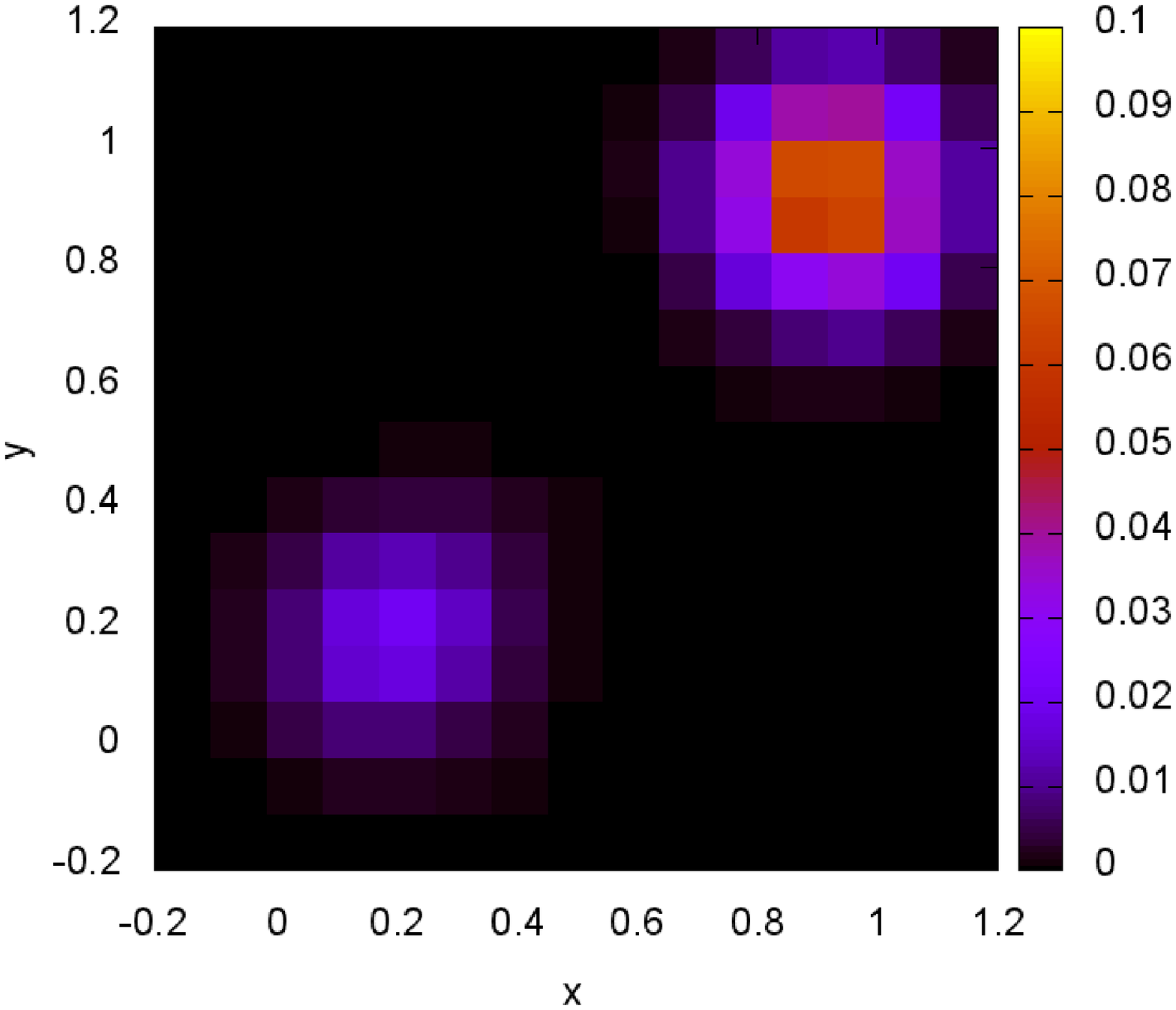}}
    \subfigure[Model 4]{\includegraphics[width=0.25\textwidth]{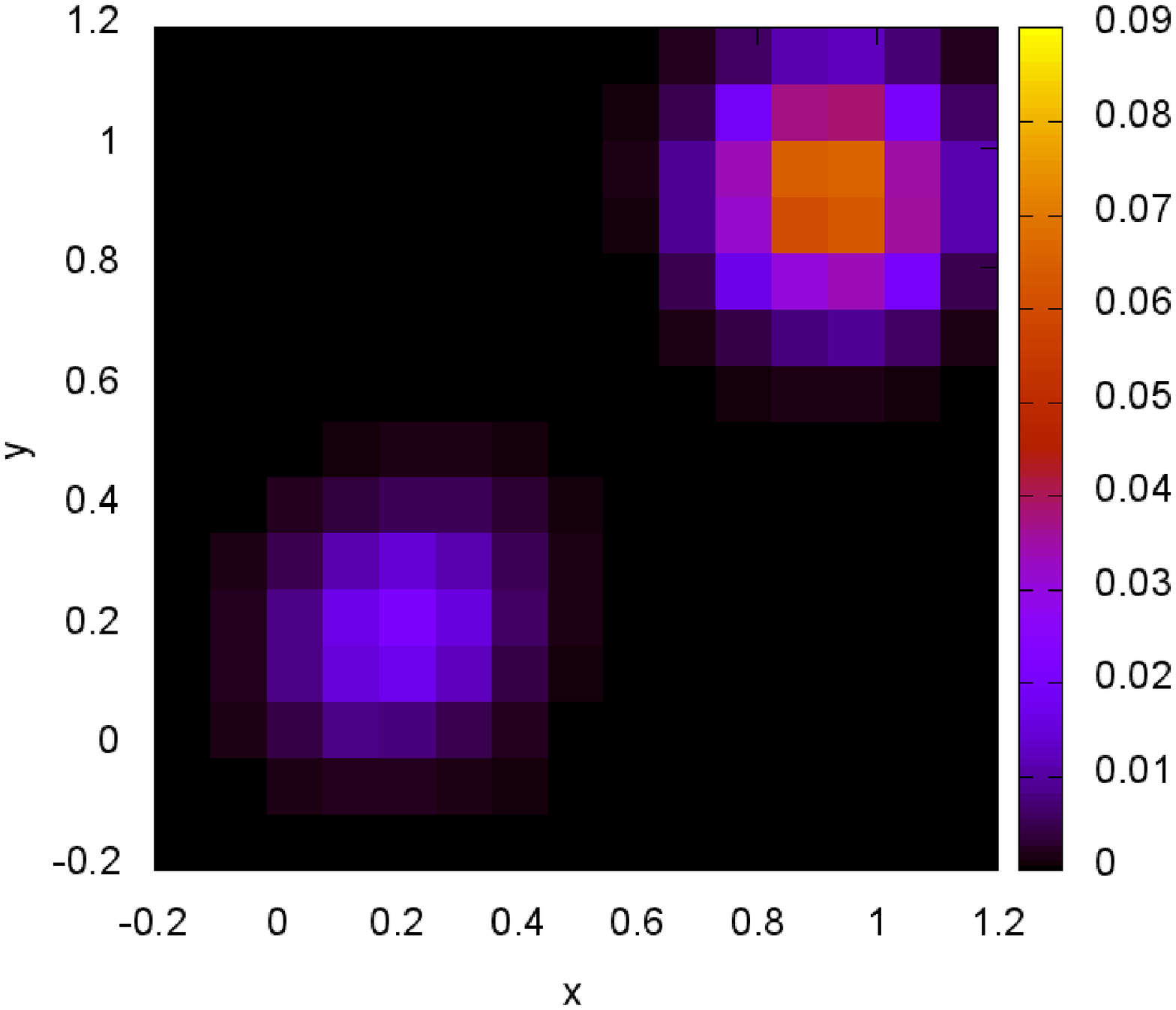}}}
  \caption{Four test models with three two-dimensional Gaussian
    distributions of width 0.1 in each dimension.  The three
    components have centres at (0.2, 0.2), (0.9, 0.9), (0.3, 0.3). The
    first and second components for all models were realised with 1000
    and 4000 counts respectively.  The third component for Models 1,
    2, 3 and 4 have have 1000, 500, 300, and 200 counts, respectively.
    Components Two and Three are barely
    distinguishable by eye, even for Models 1 and 2 and appear as an
    elongation.  The ``by eye'' differences between Models 3, 4 and
    Model 5 with zero tertiary amplitude are not easily distinguished.
    \label{fig:splats}
  }
\end{figure*}

Many problems in astronomy are mixtures of components drawn from the
same model family. Each component $j$ in the mixture is additively
combined with a weight $w_j$ such that $\sum_{j=1}^m w_j=1$.  This
allows the predefinition of some generic RJMCMC transitions that are
likely to work for a wide variety of problems.  I consider two types
of transitions.  The first is the \emph{birth} and \emph{death} of a
component.  The birth step is implemented by selecting a new component
from the prior distribution for the component parameters for the
user-specified model.  The prior for the weights is chosen here to be
a Dirichlet distribution with a single user-specified shape parameter
since each component is assumed to be indistinguishable a priori.  Assume
that the new component is born in state with $m$ components.  The new
$m+1$ weight is selected from the prior distribution for $m+1$ weights
after marginalising over $m$ of the them.  The $m$ weights in the
current state are scaled to accommodate the choice.  The death of a
component is the inverse of the birth step defined by detailed
balance.  The second type of transitions are \emph{split} and
\emph{join} transitions; again, these are inverses.  For the split
step, one selects a component at random and splits each component with
an additive or multiplicative shift as follows: $\theta_1 = \theta_0 +
\delta\theta, \theta_2 = \theta_0 - \delta\theta$, or $\theta_1 =
\theta_0(1 + \epsilon), \theta_2 = \theta_0(1 - \epsilon)$.

\begin{table}
  \caption{RJMCMC applied to the images in Figure \protect{\ref{fig:splats}}}
  \begin{threeparttable}
    \begin{tabular}{llllll}
      Model 	&$A_3/A_2$	&$A_3/A_{total}$	& $p(2)$\tnote{$\dagger$}& $p(3)$	& $p(4)$	\\ \hline
      1		& 1.0		& 0.167		& 0.0		& 0.9997	& 0.0003	\\
      2		& 0.5		& 0.091 	& 0.0		& 0.9997	& 0.0003	\\
      3		& 0.3		& 0.057		& 0.444		& 0.556		& 0.0003	\\
      4		& 0.2		& 0.038		& 0.963		& 0.036		& 0.001		\\
      5		& 0.0		& 0.0		& 0.998		& 0.002		& 0.0		\\ \hline
    \end{tabular}
    \begin{tablenotes}
    \item[$\dagger$] $p(m)$ is the probability of states in the subspace with
      $m$ components. For these simulations, $p(1)=p(5)=p(6)=0$
    \end{tablenotes}
  \end{threeparttable}
  \label{tab:splats}
\end{table}

For an example, consider a toy model for a group of galaxies where the
light from each galaxy has a normal distribution with the same width.
Each image has two well-separated components with $A_1=4000$ and
$A_2=1000$ counts. A tertiary component with one of five different
amplitudes $A_3$ is added along the line between the two, separated by
a half width.  With this choice, the secondary and tertiary components
are blended and appear as a single peak (see Fig. \ref{fig:splats} for
details).  The results of applying the reversible jump transitions
above to these four images and one with an empty tertiary component is
described in Table \ref{tab:splats}.  The total number of counts is
$A_{total}=\sum_{i=1}^3 A_i$ and $p(m)$ denotes the probability of
states in subspace $m$. The prior probability on the centre
coordinates is uniform in the range $(-0.2, 1.2)$.  The prior
probability on each subspace with $m\in[1,6]$ components is Poisson
with a mean of 1; that is, I am intentionally preferring fewer
components, but tests with Poisson means of 2 and 3 show that the
results are insensitive to this choice.  The MCMC simulations use the
tempered transitions algorithm described in Section \ref{sec:sim_temp}
with $T_{max}=8$ and 16 logarithmically-spaced temperature levels.
The Metropolis-Hastings transition probability is a uniform ball whose
widths are adjusted to achieve an acceptance rate of approximately
0.15.  The RJMCMC algorithm puts nearly all of the probability on the
two- and three-component subspaces.  The posterior distribution of
component centres are accurately constrained to the input values in
the correct subspace.  Table \ref{tab:splats} reveals a sharp
transition between preferring three to two components at an amplitude
ratio of the tertiary to secondary amplitude, $A_3/A_2$, at 0.3 (Model
3) and below.  Figure \ref{fig:splats} shows that this procedure
reflects our expectation that RJMCMC should identify three separate
components when one can do so by eye.  Of course, the subtle visual
asymmetries that allows us to do this would be obscured by noise that
would be included in a realistic model.

In summary, RJMCMC allows one to identify the number of components in
a mixture without using Bayes factors.  The simple set of transitions
used here are likely to work well for a wide variety of model families
since they depend on the mixture nature, not properties of the
underlying model families.  Unlike Bayes factors, RJMCMC simulations
may not be reused for model comparisons in light of a new model;
rather, the RJMCMC simulation must be repeated including the new
model.

\subsection{Convergence testing}
\label{sec:convergence}

The BIE provides extensible support for convergence testing.
Convergence testing has two goals: (1) to determine when the
simulation is sampling from the posterior distribution, and (2) to
determine the number of samples necessary to represent the
distribution.  Here, I address the first goal.  For multiple chains,
the work horse is the commonly used \citet{Gelman.Rubin:92} statistic.
This method compares the interchain variance to the intrachain
variance for an ensemble of chains with different initial conditions;
the similarity of the two is a necessary condition for convergence.

For single-chain algorithms, I have had good success with a diagnostic
method that assesses the convergence of both marginal and joint
posterior densities following \citet[][hereafter
GVDP]{Giakoumatos.etal:99}. This method determines confidence regions
for the posterior mean by batch sampling the chain.  As the
distribution converges, the distribution of the chain states about the
mean will approach normality owing to the central limit theorem: the
variance as a function of $1/\sqrt{N}$ for a sample size $N$ will be
linearly correlated for a converged simulation.  This approach
generalises \citet{Gelman.Rubin:92} who used the coefficient of
determination (C.O.D., the square of the Pearson's product-moment
correlation coefficient, e.g. \citealt{Press.Teukolsky.ea:92}, Section 14.5)
to assess convergence.  Moreover, GVDP use the squared ratio of the
lengths of the empirical estimated confidence intervals for the
parameters of interest as an alternative interpretation of the $R$
diagnostic.  This alternative calculation of $R$ is simpler to compute
than the original ratio of variances and is free from the assumption
of normality.

For parallel chain applications, and especially for differential
evolution (see Section \ref{sec:DE}), some chains will get stuck in regions
of anomalously low posterior probability.  Several outliers in a large
ensemble of chains can be removed without harming the simulation as
long as the chains are independent.

\subsection{Goodness-of-fit testing}

As described in Section \ref{sec:wants_needs}, model assessment is an
essential component of inference.  I have explored two approaches:
Bayesian $p$-values and Bayes factors for a non-parametric model.  The
first is easily applied but is a qualitative indicator only.  The
second has true power as a hypothesis test but is computationally
intensive.

\subsubsection{Posterior predictions}
\label{sec:postp}

Once one has successfully simulated the posterior distribution, one
may predict future data points easily.  The predicted distribution of
some future data $\data^{pred}$ after having observed the data $\data$
is
\begin{eqnarray}
  p(\data^{pred}|\data ) &=&  \int p(\data^{pred}
  , \param|\data)\,d\param \nonumber \\
  &=& \int p(\data^{pred} |\param, \data)p(\param|\data)\,d\param,
  \label{eq:pred}
\end{eqnarray}
called the \emph{posterior predictive distribution}.  In the last
integral expression in equation (\ref{eq:pred}), $p(\data^{pred}
|\param, \data)$ is the probability of observing $\data^{pred}$ given
the model parameter $\param$ and observed data set $\data$.  The
distribution $p(\param|\data)$ is the posterior distribution.  For
many problems of interest, the probability of observing some new data
given the model parameter $\param$ is independent of the original
$\data$.  In many cases, $p(\data^{pred} |\param, \data)$ will be the
standard likelihood function $p(\data^{pred} | \param)$,
simplifying equation (\ref{eq:pred}).  One may
simulate the posterior predictive distribution using an existing MCMC
sample as follows: 1) sample $m$ values of $\param$ from the
posterior; 2) for each $\param$ in a posterior set, sample a value of
$\data^{pred}$ from the likelihood $p(\data^{pred} |\param)$.  The $m$
values of $\data^{pred}$ represent samples from the posterior
predictive distribution $p(\data^{pred}|\data)$.

\subsubsection{Posterior predictive checking}
\label{sec:ppc}

One can attempt to check specific model assumptions with
\emph{posterior predictive checks} (PPC) using the posterior
predictive distribution.  The idea is simple: if the model fits,
predicted data generated under the model should look similar to the
observed data.  That is, the discrepancy measure applied to the true
data should \emph{not} lie in the tails of the predicted distribution.
If one sees some discrepancy, does it owe to an inappropriate model or
to random variance?  To answer this question \citep[following][and
references therein]{Gelman:2003}, one generates $M$ data sets,
$\data^{pred}_1,\ldots,\data^{pred}_M$ from the posterior predictive
distribution $p(\data^{pred}|\data)$.  Now one chooses some number of
test statistics $T(\data, \param)$ that measure the discrepancy
between the data and the predictive simulations.  These discrepancy
measures can depend on the data $\data$ and the parameters and
hyperparameters $\param$, which is different from standard hypothesis
testing where the test statistic only depends on the data, but not on
the parameters.  The discrepancy measures $T(\data,\param)$ need to be
chosen to investigate deviations of interest implied by the
\emph{nature} of the problem at hand. This is similar to choosing a
powerful test statistic when conducting a hypothesis test.  Any chosen
discrepancy measure must be meaningful and pertinent to the assumption
you want to test.  Examples of this approach using the BIE may be
found in \citet{lu.etal:11}.  The BIE expects a converged posterior
sample to enable such analyses.  Classes that enable automatic PPC
analyses will be part of the next BIE release.

\subsubsection{Non-parametric tests}
\label{sec:VW}

This class of goodness-of-fit tests weights a parametric null
hypothesis against a non-parametric alternative. For example, one may
wish to test the accuracy of an algorithm that has produced $n$
independent variates $\param_{1:n} = (\theta_1 , \theta_2 ,\ldots ,
\theta_n)$ intended to be normally distributed.  One would test the
null hypothesis that the true density is the normal distribution
${\cal N}(\mu,\sigma^2)$ against a diverse non-parametric class of
densities by placing a prior distribution on the null and alternative
hypotheses and calculating the Bayes factor. This leads to difficult,
high-dimensional calculations.

For the BIE, we adopt a remarkably clever method proposed by
\citet[][hereafter VW]{VerdinelliWasserman:1998} to perform a
non-parametric test without proposing alternative models directly.
The VW approach is based on the following observation: since the
cumulative distribution function for a scalar variable $\theta$,
$F(\theta)$, is strictly increasing and continuous, the inverse
$F^{-1}(u)$ for $u\in[0,1]$ is the unique real number $\theta$ such
that $F(\theta) = u$.  In the multivariate case, the inverse of the
cumulative distribution function will not be unique generally, but,
instead, one may define
\begin{equation} F^{-1}(u) = \inf_{\param\in\mathbb{R}^d} \{
F(\param) \geq u \}
\end{equation} 
for a parameter vector $\param$ of rank $d$.  Then, rather than
defining a general class of densities in $\mathbb{R}^d$ to propose the
alternative, VW consider a functional perturbation to $F$,
$G(F(\param))$ say, such that $G$ maps the unit interval onto itself.
The identity, $G(u)=u$, is the unperturbed probability
distribution. Then, the test evaluates the uniformity of the
distribution of probabilities under each hypothesis.

To construct the functional perturbation $G$,
\citeauthor{VerdinelliWasserman:1998} use a sequence of Legendre
polynomials, $\{\xi_j(\cdot), j = 1, 2, \ldots\}$, defined over the
unit interval to construct infinite exponential densities of the form
\[
g(u|\psi) = \exp\left[\sum^\infty_{j=1}\psi_j\xi_j(u) - c(\mathbf{\psi})\right]
\]
$\mathbf{\psi} = (\psi_1 , \psi_2 \ldots)$ are coefficients and
\[
c(\mathbf{\psi}) = \log\int_0^1 du
\exp\left[\sum^\infty_{j=1}\psi_j\xi_j(u)\right]
\]
is a normalising constant.  VW suggest normal priors on
$\mathbf{\psi}$: $\psi_j \sim {\cal N}(0, \tau^2/c_j^2 )$ where $\tau$
and the $c_j$ are appropriately chosen constants.  VW also specify a
hyperprior on $\tau$: $\tau\sim{\cal N}(0, w^2)$ truncated to the
positive values.  This distribution provides finite probability for
obtaining the null hypothesis near $\tau=0$ and decreases
monotonically for larger perturbations from the null, maintaining the
perturbative nature of the alternative hypothesis.

Intuitively, this development is closely related to the probabilistic
interpretation of the marginal likelihood and Bayes factors.  To see
this, consider the one-dimensional case for simplicity: let
$f(\data|\theta) = P(\theta)P(\data|\theta)$ and $F_0(\theta) =
\int^\theta_{-\infty} d\theta\,f(\data|\theta)$ and $P(\data) =
F_0(\infty)$.  If the distribution of $F_0(\theta_i)$ for
$\{\theta_i\}$ is not uniform in $[0, 1]$, one can perturb
$f(\data|\theta)$ by moving some density from a region of under
sampling to a region of over sampling and, thereby, increase
$P(\data)$.

\begin{table}
  \caption{Marginal likelihood values for Verdinelli-Wasserman tests
    described in Section \protect{\ref{sec:VW}}}
  \begin{threeparttable}
    \begin{tabular}{llll}
      Class & Model type &  $\ln P(\data|\model)$ & $B_{12}^\dagger$ \\ \hline
      VW (1) & Gaussian & $586.7^{+1.6}_{-0.01}$ &
      \multirow{2}{*}{$-1.4^{+1.6}_{-0.02}$} \\
      Fiducial (2) & Gaussian & $588.1^{+0.01}_{-0.01}$ \\ \hline
      VW (1) & Power law & $588.7^{+4.1}_{-3.5}$ &
      \multirow{2}{*}{$474.2^{+4.1}_{-3.5}$} \\
      Fiducial (2) & Power law & $114.5^{+0.01}_{-0.01}$  \\ \hline
    \end{tabular}
    \begin{tablenotes}
    \item[$\dagger$] Following the definition in Section \ref{sec:modelsel},
      $B_{12}$ denotes the odds that Model 1 is more likely than Model
      2.
    \end{tablenotes}
  \end{threeparttable}
  \label{tab:VW}
\end{table}

For an example, I apply the VW method to the following two models
defined by a two-dimensional normal distribution and by a
power-law-like distribution with unknown centres and widths in each
dimension:
\begin{eqnarray}
P_{Gauss}(x, y; \theta_x, \theta_y, \sigma_x, \sigma_y) &=&
\left(\frac{1}{2P\sigma_x\sigma_y}\right)e^{-r^2/2}, \\
P_{Power}(x, y; \theta_x, \theta_y, \sigma_x, \sigma_y, \alpha) &=&
\left(\frac{1}{2P\sigma_x\sigma_y}\right)
\frac{\alpha(1+\alpha)}{(1+r)^{2+\alpha}} \nonumber \\
\end{eqnarray}
where $r^2 \equiv x^2/\sigma_x^2 + y^2/\sigma_y^2$ and $\alpha>0$.
For the examples here, I adopt $\alpha=1$. I take the Gaussian model,
$P_{Gauss}$, or power-law model, $P_{Power}$, to be the null
hypothesis (denoted 0).  For each model, I assume the centres are
normally distributed with zero mean and unit variance and that the
variance is Weibull distributed with scale 0.03 and shape of 1.  A
1000-point data set is sampled from a two-dimensional normal
distribution centred at the origin with a root variance of each
dimension of 0.03.

I take the same model with the VW extension with $n=5$ basis functions
to be the alternative hypothesis (denoted 1) and test the support for
the two hypotheses using Bayes factors.  Although VW recommend the
choice $w=1$ for the hyperprior, I found that this tends to overfit
the extended model, not favouring the model when it is correct.
Rather I adopt $w=4$ which suppresses this tendency. I performed four
MCMC simulations with and without the VW extension and with both the
Gaussian and power-law models.  Each used the tempered differential
evolution algorithm (see Section \ref{sec:DE}) with 32 chains and
$T_{max}=32$ to obtain two million converged states.  Then, the
posterior samples were batched into groups of 250,000 states and the
marginal likelihood was computed using the algorithms described in
Section \ref{sec:evidence}.  The 90\% credible interval was computed
by bootstrap analysis from the batches.  The results are summarised in
Table \ref{tab:VW}.  One sees from the table that the true model is
preferred to the extended VW model, but only mildly.  Conversely, the
power-law model is strongly disfavoured relative to the extended model
for the normal data sample.  The marginal likelihood value for the
normally distributed data given the normal model (Row 1 in the table)
has almost the same value as the extended VW power-law model (Row 4 in
the table).  This is expected if the algorithm is doing its job of
perturbing the cumulative distribution in a way that maximises the
marginal likelihood.  As pointed out by VW, the extended model
provides a estimator of the true posterior density.  The agreement of
these two values suggests that the five basis functions provide
sufficient variation to reproduce the true value and that the
ten-dimensional numerical evaluation of the marginal likelihood
integral using the methods described in Section \ref{sec:evidence} is
sufficiently accurate.

\section{BIE: technical overview}
\label{sec:software}

The BIE is a general-purpose system for simulating Bayesian posterior
distributions for statistical inference and has been stress tested
using high-dimensional models.  As described in the previous sections,
the inference approach uses the Bayesian framework enabled by Monte
Carlo Markov chain (MCMC) techniques.  The software is parallel and
multi-threaded and should run in any environment that supports POSIX
threads and the widely-implemented Message Passing Interface (MPI, see
{\tt http://www.mpi-forum.org}).  The package is written in C++ and
developed on the GNU/Linux platform but it should port to any GNU
platform.

BIE at its core is a software library of interoperable components
necessary for performing Bayesian computation.  The BIE classes are
available as both C++ libraries and as a stand-alone system with an
integrated command-line interface.  The command-line interface is well
tested and is favoured by most users so far.  A user does not need to
be an expert or even an MPI programmer to use the system; the simple
user interface is similar to MatLab or Gnuplot.  In addition to the
engine itself, the BIE package includes a number of stand-alone
programs for viewing and analysing output from the BIE, testing the
convergence of a simulation, manipulating the simulation output, and
computing the marginal likelihood using the algorithms described in
Section \ref{sec:evidence}.  A future release will include wrappers for
Python.

\subsection{Software architecture}
\label{sec:arch}

As in GIMP Toolkit (GTK+) and the Visualisation Toolkit (VTK), the BIE
uses C++ to facilitate implementing an object-oriented design.
Object-oriented programming enforces an intimate relationship between
the data and procedures: the procedures are responsible for
manipulating the data to reflect its behaviour.  The software objects
(classes in C++) represent real-world probability distributions,
mathematical operators and algorithms, and this presents a natural
interface and set of interobject relationships to the user and the
developer.  Programs that want to manipulate an object have to be
concerned only about which messages this object understands, and do
not have to worry about how these tasks are achieved nor the internal
structure of the object.

Another powerful feature of object-oriented programming is
\emph{inheritance} (derived classes in C++).  The derived class
inherits the properties of its base class and also adds its own data
and routines.  This structure makes it is easy to make minor changes
in the data representation or the procedures.  Changes inside a class
do not affect any other part of a program, since the only public
interface that the external world has to a class is through the use of
methods.  This facilitates adding new features or responding to
changing operating environments by introducing a few new objects and
modifying some existing ones.  These features encourage extensibility
through the reuse of commonly used structures and innovation by
allowing the user to connect components in new and possibly unforeseen
ways.  In addition, this facilitates combining concurrently developed
software contributions from scientists interested in specific models
or data types and MCMC algorithms.

Motivated by handling large amounts of survey data with the subsequent
possible need to investigate the appropriate simulation for a variety
of models or hypotheses, the BIE separates the computation into a
collection of subsystems:
\begin{enumerate}
\item Data input and output
\item Data distribution and spatial location
\item Markov chain simulation
\item Likelihood computation
\item Model and hypothesis definition
\end{enumerate}
Each of these can be specified independently and easily mixed and
matched in our object-oriented architecture.  The cooperative
development enabled by this architecture is similar to that behind the
open-source movement, and this project is a testament to its success
in an interdisciplinary scientific collaboration. The BIE use SVN
version management (autoconf, automake), GNU coding standards, and
DejaGNU regression testing to aid in portability.  Moreover, this same
social model will extend to remote collaborative efforts as the BIE
project matures.

\subsection{Persistence system}
\label{sec:persist}

The researcher needs to be able to stop, restart, and possibly refocus
inferential computations for both technical and scientific reasons.
The BIE was designed with these scenarios in mind.  The BIE's
persistence system is built on top of the BOOST
(\url{http://www.boost.org}) serialisation library.  The BIE classes
inherit from a base serialisation class that provides the key
serialisation members and a simple mnemonic scheme to mark persistent
data in newly developed classes.  The serialisation-based persistence
system avoids irrevocably modifying data sets or files. Rather it
views the computation as a series of functions, each accepting one or
more input data sets and producing one or more new output data
sets. The system records and time stamps these computations and the
relationships between inputs and outputs in an archive research log, so
that one can always go back and determine the origin of data and how
it was processed. The most common use of BIE persistence to date is
checkpointing and recovery, Checkpointing guards against loss of
computation by saving intermediate data to support recovery in the
middle of long-running computational steps; and it allows one to
``freeze'' or ``shelve'' a computation and pick it up later. It also
provides the basic support needed to interrupt a computation, do some
reconfiguring, and resume, as when machines need to be added to or
removed from a cluster, etc.

\subsection{Extensibility}

BIE is designed to be extensible.  The user may define new classes
for any aspect of the MCMC simulation, such as MCMC algorithms,
convergence tests, prior distributions, data types, and likelihood
functions.  The code tree includes a {\tt Projects} directory
that is automatically compiled into any local build that may contain
any locally added functionality.  For example, both the BIE-SAM and
Galphat were derived from a base class specifically for user-defined
likelihood functions.

The source tree is available for download from
\url{http://www.astro.umass.edu/bie}.  The package includes Debian and
Ubuntu package management scripts so that local {\tt.deb} packages may
be built.  Users have had success building other modern Linux
distributions.

\vfill
\label{lastpage}
\eject % force out this label

\end{document}